\pdfoutput=1

\documentclass[11pt,twoside,a4paper,cmspaper,final,collab]{cms-tdr}
\def\svnVersion{458026P}\def\cmsCernNoTag{CERN-EP-2017-120}\def\cmsCernDate{\today}\def\cmsMessage{Published in the Journal of High Energy Physics as \href{http://dx.doi.org/10.1007/JHEP04(2018)060}{\doi{10.1007/JHEP04(2018)060.}}}

\begin{document}\cmsNoteHeader{TOP-16-007}

\hyphenation{had-ron-i-za-tion}
\hyphenation{cal-or-i-me-ter}
\hyphenation{de-vices}
\RCS$HeadURL: svn+ssh://svn.cern.ch/reps/tdr2/papers/TOP-16-007/trunk/TOP-16-007.tex $
\RCS$Id: TOP-16-007.tex 363267 2016-08-03 20:06:37Z linacre $
\newlength\cmsFigWidth
\setlength\cmsFigWidth{0.48\textwidth}

\newcommand{\ptlep}{\ensuremath{\pt^{\smash[b]{\,\text{lep}}}}\xspace}
\newcommand{\ptjet}{\ensuremath{\pt^{\smash[b]{\,\text{jet}}}}\xspace}
\newcommand{\ptt}{\ensuremath{\pt^{\PQt}}\xspace}
\newcommand{\yt}{\ensuremath{y^{\PQt}}\xspace}
\newcommand{\pttt}{\ensuremath{p_{\mathrm{T}}^{\smash[b]{\ttbar}}}\xspace}
\newcommand{\ytt}{\ensuremath{y^{\ttbar}}\xspace}
\newcommand{\mtt}{\ensuremath{M^{\ttbar}}\xspace}
\newcommand{\dphtt}{\ensuremath{\Delta \phi^{\ttbar}}\xspace}

\newcommand{\cmsSF}[1]{{\small \textsf{#1}}\xspace}
\newcommand{\cmsRM}[1]{{\small \textrm{#1}}\xspace}
\newcommand\aMCatNLO{{\cmsRM{MG5}}\_a\MCATNLO}
\newcommand\FXFX{\cmsRM{[FXFX]}}
\newcommand\MLM{\cmsRM{[MLM]}}
\newcommand{\x}{\ensuremath{\phantom{0}}}
\newcommand{\y}{\ensuremath{\phantom{.}}}
\newcommand{\z}{\ensuremath{\phantom{<}}}
\newcommand{\w}{\ensuremath{\phantom{-}}}

\cmsNoteHeader{TOP-16-007}

\title{Measurement of normalized differential \texorpdfstring{$\ttbar$}{t-tbar} cross sections in the dilepton channel from pp collisions at \texorpdfstring{$\sqrt{s}=13\TeV$}{sqrt(s) = 13 TeV}}

\date{\today}

\abstract{
Normalized differential cross sections for top quark pair production are measured in the dilepton ($\Pe^{+}\Pe^{-}$, $\PGm^{+}\PGm^{-}$, and $\PGm^{\mp}\Pe^{\pm}$) decay channels in proton-proton collisions at a center-of-mass energy of 13\TeV.  The measurements are performed with data corresponding to an integrated luminosity of 2.1\fbinv using the CMS detector at the LHC. The cross sections are measured differentially as a function of the kinematic properties of the leptons, jets from bottom quark hadronization, top quarks, and top quark pairs at the particle and parton levels. The results are compared to several Monte Carlo generators that implement calculations up to next-to-leading order in perturbative quantum chromodynamics interfaced with parton showering, and also to fixed-order theoretical calculations of top quark pair production up to next-to-next-to-leading order.
}

\hypersetup{%
pdfauthor={CMS Collaboration},%
pdftitle={Measurement of normalized differential ttbar cross sections in the dilepton channel from pp collisions at sqrt(s) = 13 TeV},%
pdfsubject={CMS},%
pdfkeywords={CMS, top quark, differential cross section}}

\maketitle

\section{Introduction}

The measurement of \ttbar differential cross sections can provide a test of perturbative quantum chromodynamic (QCD) calculations and also improve the knowledge of parton distribution functions (PDFs)~\cite{Czakon:2013tha}. Previous measurements of differential cross sections for \ttbar production have been performed in proton-proton (pp) collisions at the CERN LHC at center-of-mass energies of  7~\cite{Chatrchyan:2012saa, Aad:2015eia} and 8\TeV~\cite{Khachatryan:2015oqa,Aad:2015mbv,Khachatryan:2016oou, Khachatryan:2016gxp, Khachatryan:2015fwh,Khachatryan:2015mva,Aaboud:2016omn,Aaboud:2016iot, Sirunyan:2254647}.  The  dilepton (electron or muon) final state of the ${\ttbar}$ decay helps in the suppression of background events. This paper presents the first CMS measurement at $\sqrt{s} =  13\TeV$ in the dilepton decay final state and includes the same-flavor lepton channels ($\Pe^{+}\Pe^{-}$ and $\PGm^{+}\PGm^{-}$), using data corresponding to an integrated luminosity of 2.1\fbinv. The statistical precision of the measurements is improved by the increased data sample from including the same-flavor lepton channels. The data were recorded by the CMS experiment at the LHC in 2015, and this measurement complements other recent measurements that have been reported in a different decay channel~\cite{Khachatryan:2016mnb} and by a different experiment~\cite{Aaboud:2016xii, Aaboud:2016syx}.

The \ttbar differential cross section measurements are performed at the particle and parton levels. Particle-level measurements use final-state kinematic observables that are experimentally measurable and theoretically well defined. Corrections are limited mainly to detector effects that can be determined experimentally. The particle-level measurements are designed to have minimal model dependencies. The visible differential cross section is defined for a phase space within the acceptance of the experiment. Large extrapolations into inaccessible phase-space regions are thus avoided in particle-level differential cross section measurements. In contrast, the parton-level measurement of the top quark pair production cross sections is performed in the full phase space. This facilitates comparisons to predictions in perturbative QCD.

The normalized \ttbar differential cross sections are measured as a function of the kinematic properties of the \ttbar system, the top quarks and the top quark decay products, which include the jets coming from the hadronization of bottom quarks and the leptons. The particle-level measurements are performed with respect to the transverse momentum of the leptons and of the jets. The cross sections as a function of the invariant mass and rapidity of the \ttbar system are also measured to help in understanding the PDFs. The angular difference in the transverse plane between the top and anti-top quarks is provided to compare to predictions of new physics beyond the standard model~\cite{bib:choi}. In addition, the normalized \ttbar cross sections are measured as a function of the transverse momenta of the top quark and of the top quark pair.

\section{The CMS detector and simulation}
\subsection{The CMS detector}

The central feature of the CMS apparatus is a superconducting solenoid of 6\unit{m} internal diameter, providing a magnetic field of 3.8\unit{T}. The solenoid volume encases the silicon pixel and strip tracker, a lead tungstate crystal electromagnetic calorimeter, and a brass and scintillator hadron calorimeter, each composed of a barrel and two endcap sections. Forward calorimeters extend the pseudorapidity ($\eta$) coverage provided by the barrel and endcap detectors. Muons are detected in gas-ionization chambers embedded in the steel flux-return yoke outside the solenoid. A more detailed description of the CMS detector, together with a definition of the coordinate system used and the relevant kinematic variables, can be found in Ref.~\cite{Chatrchyan:2008zzk}. The particle-flow (PF) algorithm~\cite{Sirunyan:2017ulk} is used to reconstruct objects in the event, combining information from all the CMS subdetectors. The missing transverse momentum vector (\ptvecmiss) is defined as the projection onto the plane perpendicular to the beam axis of the negative vector sum of the momenta of all PF candidates in an event~\cite{Chatrchyan:2011tn}. Its magnitude is referred to as \ptmiss.

\subsection{Signal and background simulation}

Monte Carlo (MC) techniques are used to simulate the \ttbar signal and the background processes. We use the \POWHEG (v2)~\cite{Nason:2004rx,Frixione:2007vw,Alioli:2010xd,Campbell:2014kua} generator to model the nominal \ttbar signal at next-to-leading order (NLO). In order to simulate \ttbar events with additional partons, \MADGRAPH{}5\_a\MCATNLO (v2.2.2)~\cite{Alwall:2014hca} (\aMCatNLO) is used, which includes both leading-order (LO) and NLO matrix elements (MEs). Parton shower (PS) simulation is performed with \PYTHIA{}8 (v8.205)~\cite{Sjostrand:2007gs}, using the tune CUETP8M1~\cite{Skands:2014pea} to model the underlying event. Up to two partons in addition to the \ttbar pair are calculated at NLO and combined with the \PYTHIA{}8 PS simulation using the FXFX~\cite{Frederix2012} algorithm, denoted as {\aMCatNLO}+\PYTHIA{}8\FXFX.  Up to three partons are considered at LO and combined with the \PYTHIA{}8 PS simulation using the MLM~\cite{Alwall:2007fs} algorithm, denoted as {\aMCatNLO}+\PYTHIA{}8\MLM. The data are also compared to predictions obtained with \POWHEG samples interfaced with \HERWIGpp~\cite{Bahr:2008pv} (v 2.7.1) using the tune EE5C~\cite{Seymour:2013qka}. The signal samples are simulated assuming a top quark mass of 172.5\GeV and normalized to the inclusive cross section calculated at NNLO precision with next-to-next-to-leading-logarithmic (NNLL) accuracy~\cite{Czakon:2011xx}.

For the simulation of \PW~boson production and the Drell--Yan process, the \aMCatNLO generator is used, and the samples are normalized to the cross sections calculated at NNLO~\cite{Li:2012wna}. The $t$-channel single top quark production in the tW channel is simulated with the \POWHEG generator based on the five-flavor scheme~\cite{Frixione:2008yi, Re:2010bp}, and normalized to the cross sections calculated at NNLO~\cite{Kidonakis:2012rm}. Diboson samples ($\PW\PW$, $\PW\PZ$, and $\PZ\PZ$) are simulated at LO using \PYTHIA{}8, and normalized to the cross section calculated at NNLO for the $\PW\PW$ sample~\cite{PhysRevLett.113.212001} and NLO for the $\PW\PZ$ and $\PZ\PZ$ samples~\cite{Campbell:2010ff}.

The detector response to the final-state particles is simulated using \GEANTfour~\cite{Allison:2006ve,Agostinelli:2002hh}. Additional pp collisions in the same or nearby beam crossings (pileup) are also simulated with \PYTHIA{}8 and superimposed on the hard-scattering events using a pileup multiplicity distribution that reflects that of the analyzed data.
Simulated events are reconstructed and analyzed with the same software used to process the data.

\section{Object and event selection}
\label{sec:EventSel}

The dilepton final state of the ${\ttbar}$ decay consists of two leptons (electrons or muons), at least two jets, and \ptmiss from the two neutrinos. Events are selected using dilepton triggers with asymmetric \pt thresholds. The low transverse momentum (\pt) threshold is  8\GeV for the muon and 12\GeV for the electron, and the high-\pt threshold is 17\GeV for both muon and electron. The trigger efficiency is measured in data using triggers based on \ptmiss~\cite{Khachatryan:2016kzg}.

The reconstructed and selected muons~\cite{Chatrchyan:2012xi} and electrons~\cite{Khachatryan:2015hwa} are required to have $\pt > 20\GeV$ and $\abs{\eta}<2.4$. Since the primary leptons that originated from top quark decays are expected to be isolated, an isolation criterion is placed on each lepton to reduce the rate of secondary leptons from non-top hadronic decays.  A relative isolation parameter is used, which is calculated as the sum of the \pt of charged and neutral hadrons and photons in a cone of angular radius $\Delta R = \sqrt{\smash[b]{(\Delta \phi)^2 +(\Delta \eta)^2 )}}$ around the direction of the lepton, divided by the lepton \pt, where $\Delta \phi$ and $\Delta \eta$ are the azimuthal and pseudorapidity differences, respectively, between the directions of the lepton and the other particle. Any mismodeling of the lepton selection in the simulation is accounted for by applying corrections derived using a ``tag-and-probe'' technique based on control regions in data~\cite{Khachatryan:2015uqb}.

Jets are reconstructed using PF candidates as inputs to the anti-\kt jet clustering algorithm~\cite{Cacciari:2011ma,Cacciari:2008gp}, with $ \Delta R = 0.4 $. The momenta of jets are corrected to account for effects from pileup, as well as nonuniformity and nonlinearity of the detector. For the data, energy corrections are also applied to correct the detector response~\cite{Khachatryan:2016kdb}. We select jets with $ \pt > 30\GeV $ and $ \abs{\eta} < 2.4 $ that pass identification criteria designed to reject noise in the calorimeters.

Jets from the hadronization of \PQb~quarks (\PQb~jets) are identified by the combined secondary vertex \PQb~tagging algorithm~\cite{Chatrchyan:2012jua}. The jets are selected using a loose working point~\cite{CMS-PAS-BTV-15-001}, corresponding to an efficiency of about $80\%$ and a light-flavor jet rejection probability of $85\%$. The \PQb~tagging efficiency in the simulation is corrected to be consistent with that in data.

Events are required to have exactly two oppositely charged leptons with the invariant mass of the dilepton system $M_{\ell^{+}\ell^{-}}\, > 20\GeV$, and two or more jets, at least one of which has to be identified as a \PQb~jet.
For the same-flavor lepton channels ($\Pe\Pe$ and $\Pgm\Pgm$), additional selection criteria are applied to reject events from Drell--Yan production: $ \ptmiss > 40\GeV$ and $|M_{\ell^{+}\ell^{-}}\,- M_{\PZ}| > 15\GeV$, where $M_{\PZ}$  is the $\PZ$ boson mass~\cite{PDG2016}. The selected numbers of events after the selection are listed in Table~\ref{tab:event_table}.

\begin{table}[htb]
\centering
\topcaption{The expected and observed numbers of events after selection are listed in the second column. The third column shows the numbers of reconstructed \ttbar events.}
\label{tab:event_table}
\renewcommand{\arraystretch}{1.3}
{
\begin{tabular}{l |c | c}
 Dilepton  &  Selected  & Reconstructed \ttbar system  \\ \hline
\ttbar-signal &     11565 $\pm$  14.19    &   10611  $\pm$  13.61  \\ \hline
\ttbar-others &  6060 $\pm$  10.28   &     4856  $\pm$   9.24 \\
Single top  &      869 $\pm$   7.93    &     540 $\pm$   6.32   \\
Dibonson & $73 \pm   3.91$    & $      39 \pm   2.87$  \\
$\PW+\text{jets}$ &  $      23 \pm  10.84$    & $      36 \pm  16.93$  \\
$\Z/\gamma* \to \ell^{+}\ell^{-}$ & $     507 \pm  12.86$    & $     324 \pm  10.75$  \\ \hline
MC total  & $   19100 \pm  25.85$    & $   16409 \pm  26.85$   \\ \hline
Data &    18891              &   16325    \\
\end{tabular}
}
\end{table}

\section{Signal definition}
\label{sec:signal}

The measurements of normalized \ttbar differential cross sections are performed at both particle and parton levels as a function of kinematic observables, defined at the generator level. The particle-level top quark is defined at the generator level using the procedure described below. This approach avoids theoretical uncertainties in the measurements due to the different calculations within each generator, and leads to results that are largely independent of the generator implementation and tuning. Top quarks are reconstructed in the simulation starting from the final-state particles with a mean lifetime greater than 30\unit{ps} at the generator level, as summarized in Table~\ref{tab:phase}.

Leptons are ``dressed'', which means that leptons are defined using the anti-\kt clustering algorithm~\cite{Cacciari:2011ma,Cacciari:2008gp} with $\Delta R = 0.1 $ to account for final-state radiated photons. To avoid the ambiguity of additional leptons at the generator level, the clustering is applied to electrons, muons, and photons not from hadron decays. Events with leptons associated with $\tau$ lepton decays are treated as background. Leptons are required to satisfy the same acceptance requirements as imposed on the reconstructed objects described in Section~\ref{sec:EventSel}, \ie, $\pt > 20\GeV$ and $ \abs{\eta} < 2.4 $.

The generator-level jets are clustered using the anti-\kt algorithm with $ \Delta R = 0.4 $.  The clustering is applied to all final-state particles except neutrinos and particles already included in the dressed-lepton definition. Jets are required to have $ \pt > 30\GeV $ and $\abs{\eta}< 2.4$ to be consistent with the reconstructed-object selection.
To identify the bottom quark flavor of the jet, the ghost-B hadron technique~\cite{Khachatryan:2016mnb} is used in which short-lifetime B hadrons are included in the jet clustering after scaling down their momentum to be negligible. A jet is identified as a \PQb~jet if it contains any B~hadrons among its constituents.

\begin{table}[htb]
\centering
\topcaption{Summary of the object definitions at the particle level.}\label{tab:phase}
\begin{tabular}{c|l|c}
Object & \multicolumn{1}{c|}{Definition} & Selection criteria \\
\hline
Neutrino & neutrinos not from hadron decays & none \\
\hline
\multirow{3}{*}{Dressed lepton} & anti-\kt algorithm with $ \Delta R = 0.1 $           & \\
& using electrons, muons, and photons  & $\pt > 20\GeV$, $ \abs{\eta} < 2.4 $ \\
& not from hadron decays               & \\
\hline
\multirow{4}{*}{\PQb~quark jet} & anti-\kt algorithm with $ \Delta R = 0.4 $               &  \\
& using all particles and ghost-B hadrons & $ \pt > 30\GeV $, $ \abs{\eta} < 2.4 $ \\
& not including any neutrinos             & with ghost-B hadrons \\
& nor particles used in dressed leptons   & \\
\end{tabular}
\end{table}

A $\PW$~boson at the particle level is defined by combining a dressed lepton and a neutrino. In each event, a pair of particle-level $\PW$~bosons is chosen among the possible combinations such that the sum of the absolute values of the invariant mass differences with respect to the $\PW$~boson mass is minimal~\cite{PDG2016}. Similarly, a top quark at the particle level is defined by combining a particle-level $\PW$~boson and a \PQb~jet. The combination of a $\PW$~boson and a \PQb~jet with the minimum invariant mass difference from the correct top quark mass~\cite{PDG2016} is selected. Events are considered to be in the visible phase space if they contain a pair of particle-level top quarks, constructed from neutrinos, dressed leptons, and \PQb~jets. Simulated dilepton events that are not in the visible phase space are considered as background and combined with the non-dilepton \ttbar decay background contribution, subsequently denoted as \ttbar-others.

In addition, the top quark and \ttbar system observables are defined before the top quark decays into a bottom quark and a $\PW$~boson and after QCD radiation, which we refer to as the parton level. The \ttbar system at the parton level is calculated in the generator at NLO. The normalized differential cross sections at the parton level are derived by extrapolating the measurements into the full phase space, which includes the experimentally inaccessible regions, such as at high rapidity and low transverse momentum of the leptons and jets.

\section{Reconstruction of the \texorpdfstring{$\ttbar$}{t-tbar} system}
\label{sec:TopReco}

The top quark reconstruction method is adopted from the recent CMS measurement of the differential \ttbar cross section~\cite{Khachatryan:2015oqa}. In the dilepton channel, the reconstruction of the neutrino and antineutrino is crucial in measuring the top quark kinematic observables. Using an analytical approach~\cite{bib:Sonnenschein,Dalitz:1992np}, the six unknown neutrino degrees of freedom are constrained by the two measured components of \ptvecmiss and the assumed invariant masses of both the \PW~boson and top quark. The efficiency for finding a physical solution depends on the detector resolution, which is accounted for by reconstructing the \ttbar system in both the MC simulation and data with 100 trials, using random modifications of the measured leptons and b~jets within their resolution functions. The efficiency for finding a physical solution to the kinematic reconstruction is approximately 90\%, as determined from simulation and data. The numbers of events remaining after reconstructing the ttbar system are listed in Table~\ref{tab:event_table}.

In each trial, the solution with the minimum invariant mass of the \ttbar system is selected, and a weight is calculated based on the expected invariant mass distribution of the lepton and \PQb~jet pairs ($M_{\ell \PQb}$) at generator level. The lepton and \PQb~jet pairs with the maximum sum of weights are chosen for the final solution of the \ttbar system, and the reconstructed neutrino momentum is taken from the weighted average over the trials.

\begin{figure}
\centering
\includegraphics[width=\cmsFigWidth]{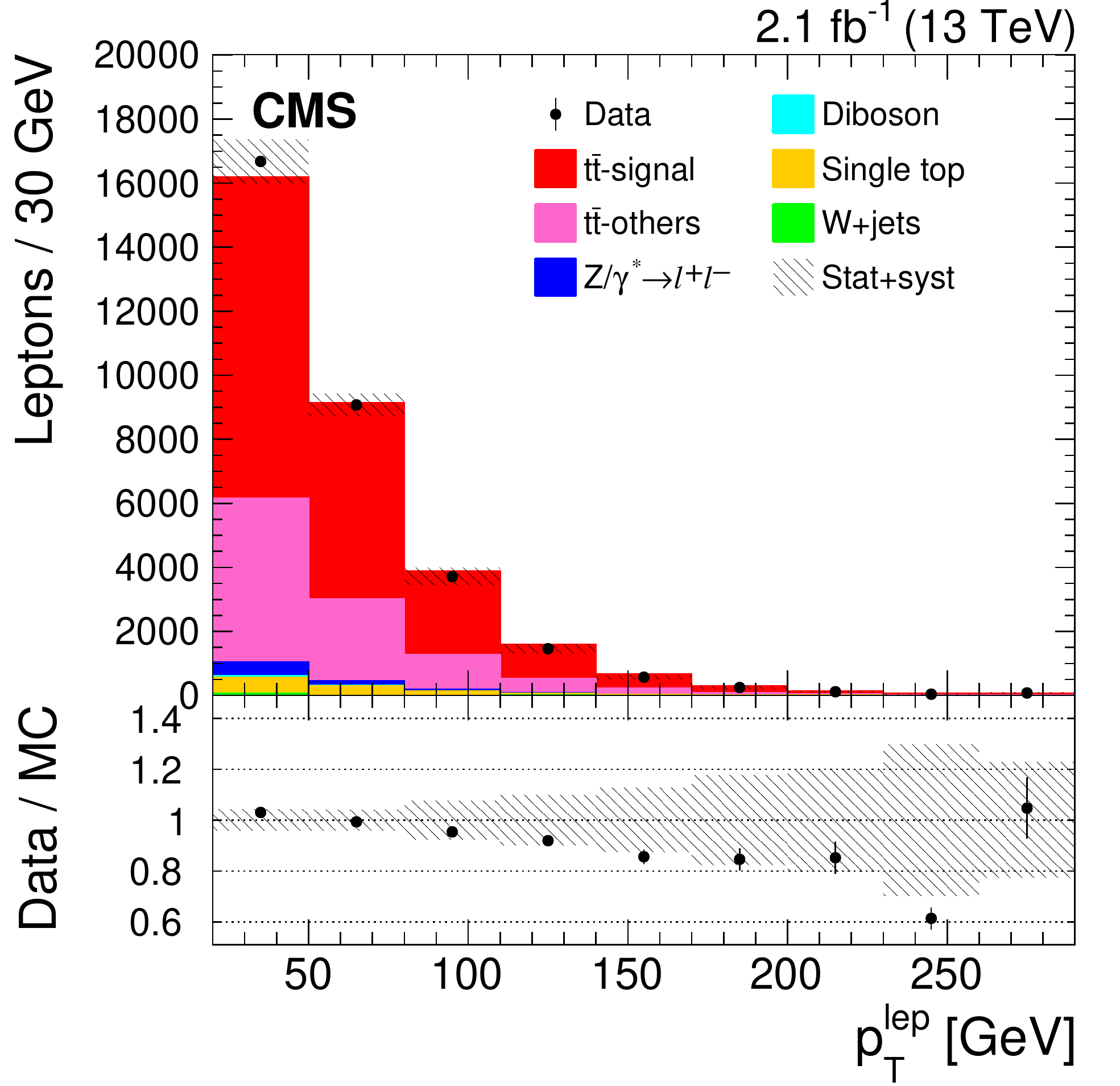}
\includegraphics[width=\cmsFigWidth]{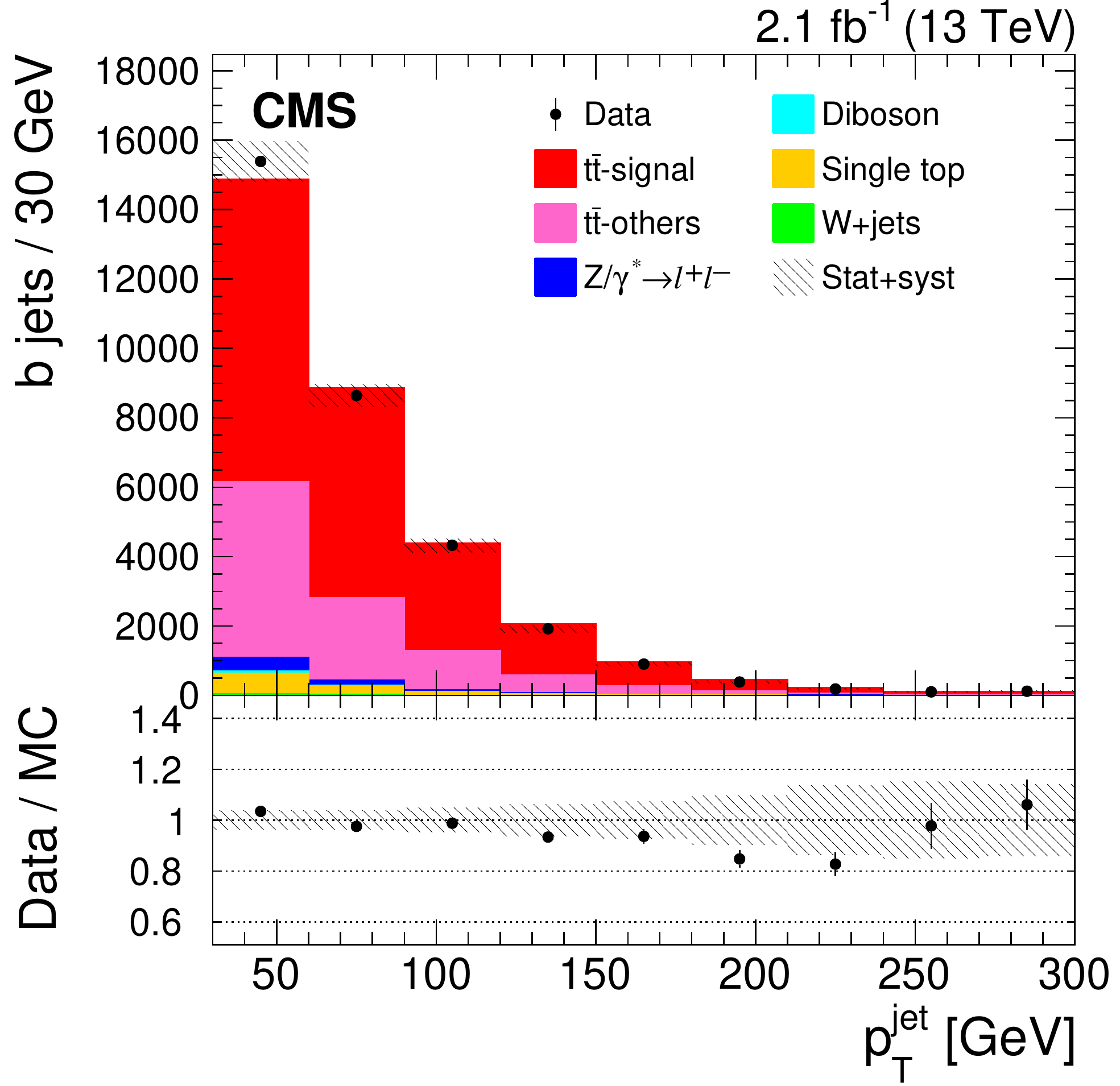}
\\ \vspace{0.2cm}
\includegraphics[width=\cmsFigWidth]{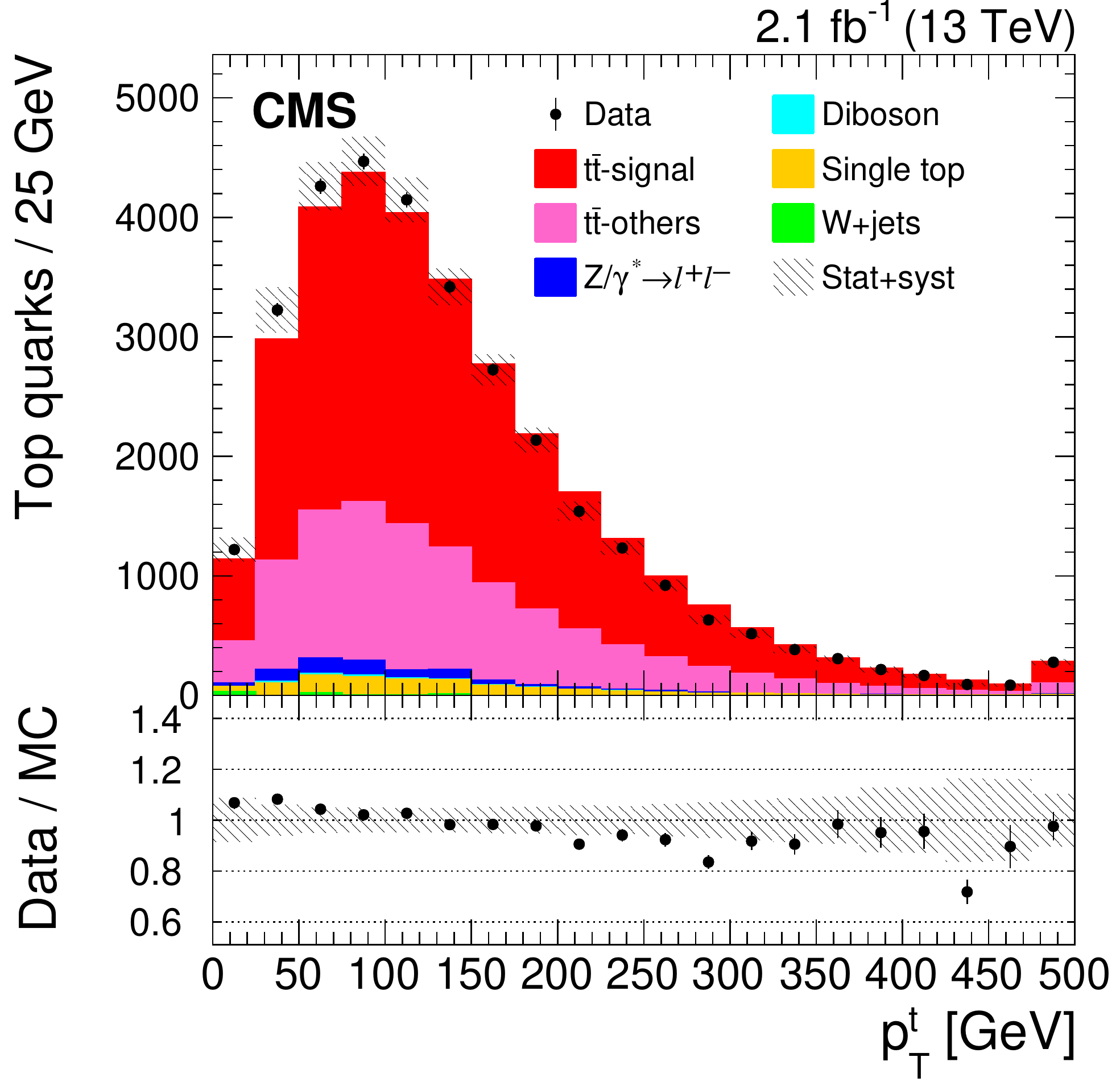}
\includegraphics[width=\cmsFigWidth]{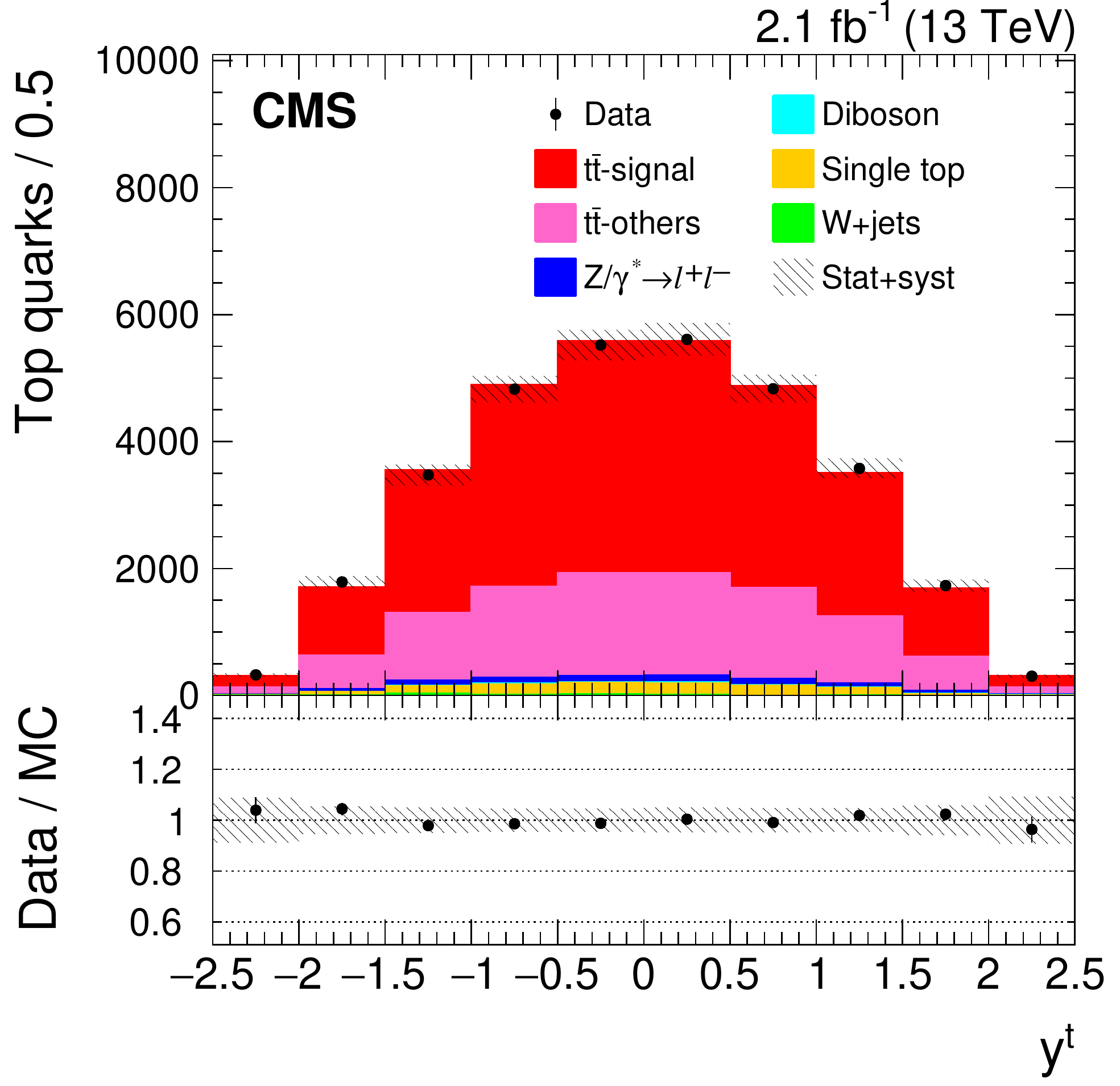}
\caption{Reconstructed $\ptlep$ (upper left), $\ptjet$ (upper right), $\ptt$ (lower left), and $\yt$ (lower right) distributions from data (points) and from MC simulation (shaded histograms). The signal definition for particle level is considered to distinguish \ttbar-signal and \ttbar-others. All corrections described in the text are applied to the simulation. The last bin includes the overflow events. The uncertainties shown by the vertical bars on the data points are statistical only while the hatched band shows the combined statistical and systematic uncertainties added in quadrature. The lower panels display the ratios of the data to the MC prediction. }
\label{fig:dist1}
\vspace{10pt}
\end{figure}

\begin{figure}
\centering
\includegraphics[width=\cmsFigWidth]{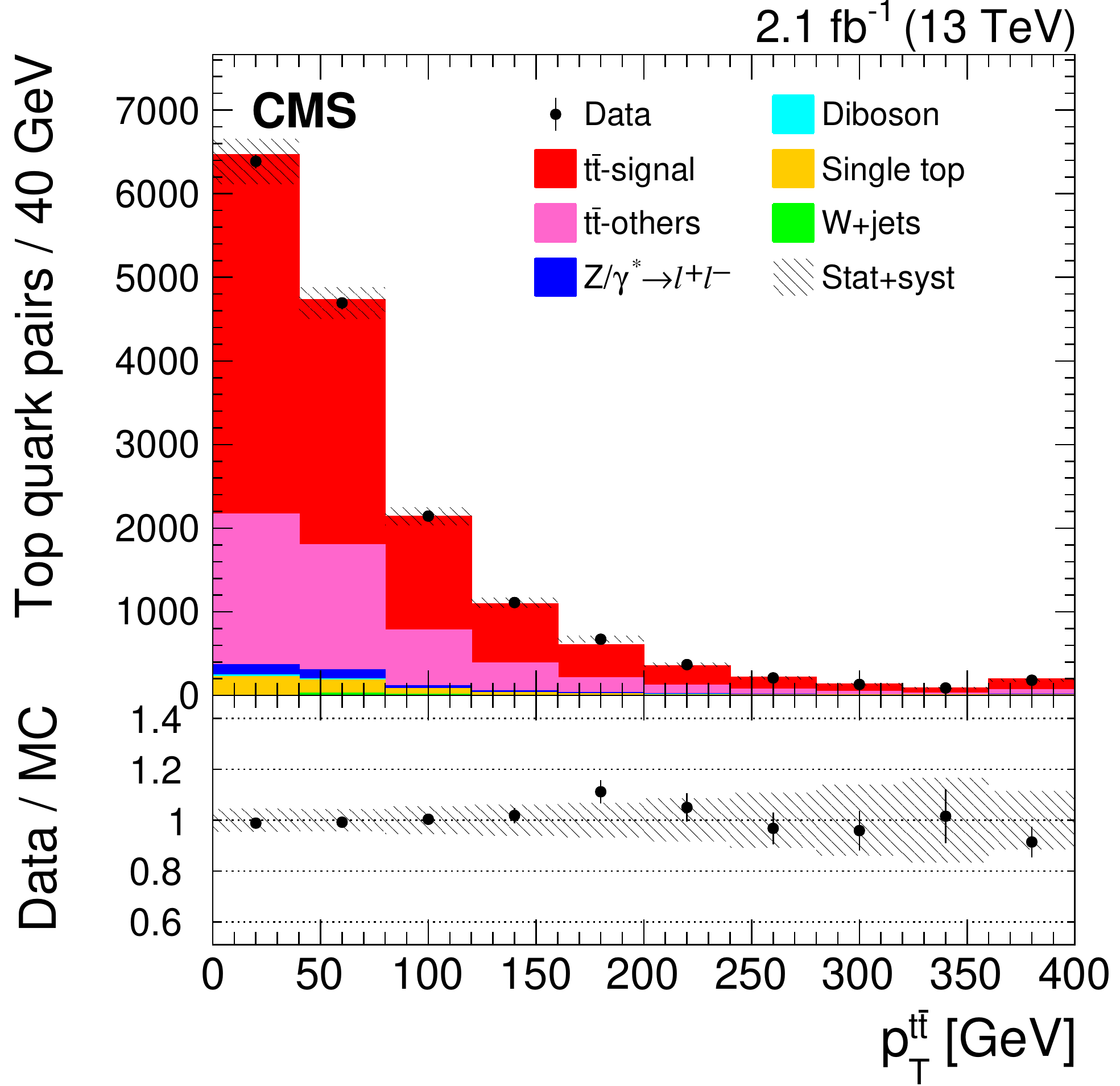}
\includegraphics[width=\cmsFigWidth]{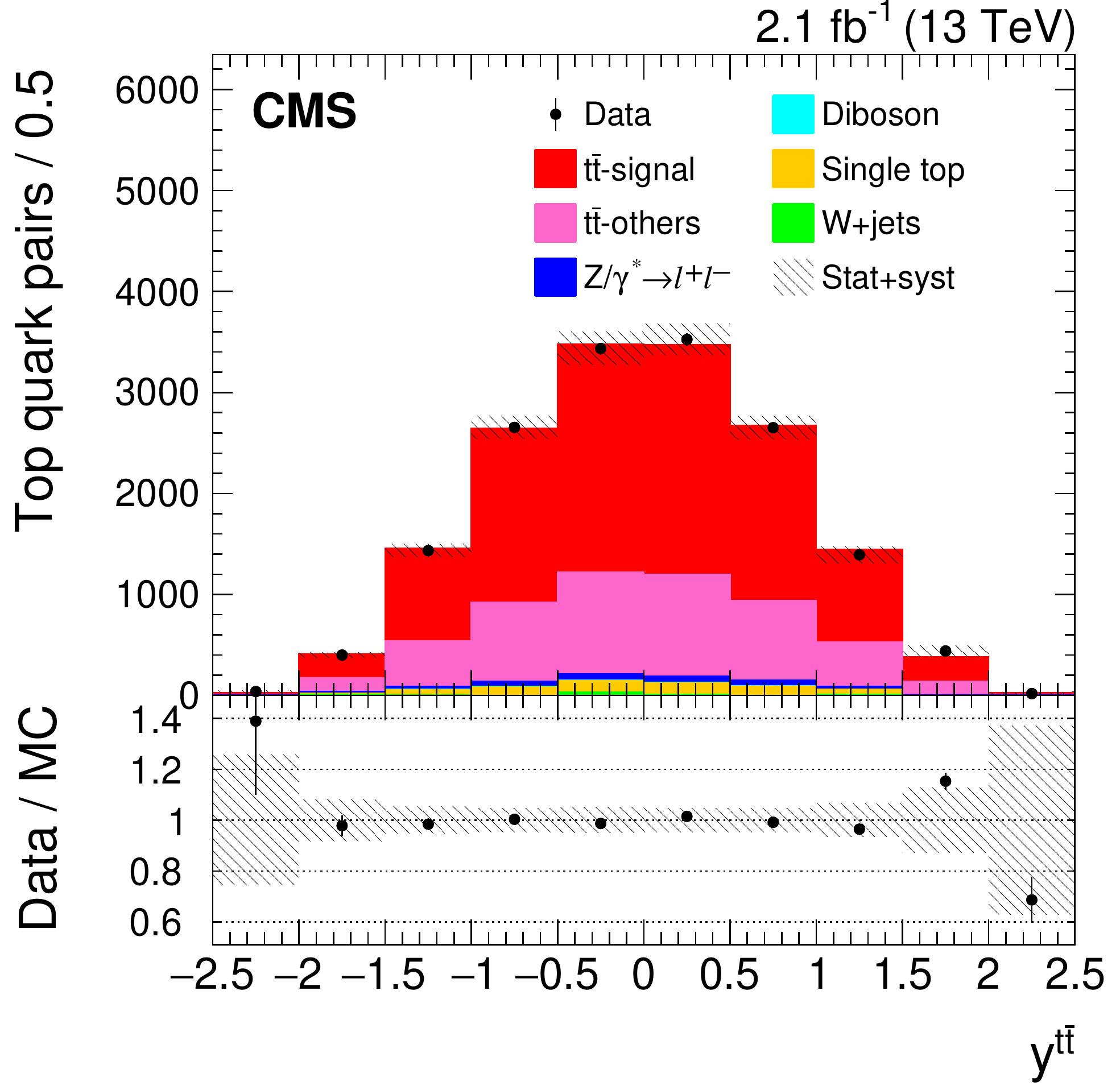}
\\ \vspace{0.2cm}
\includegraphics[width=\cmsFigWidth]{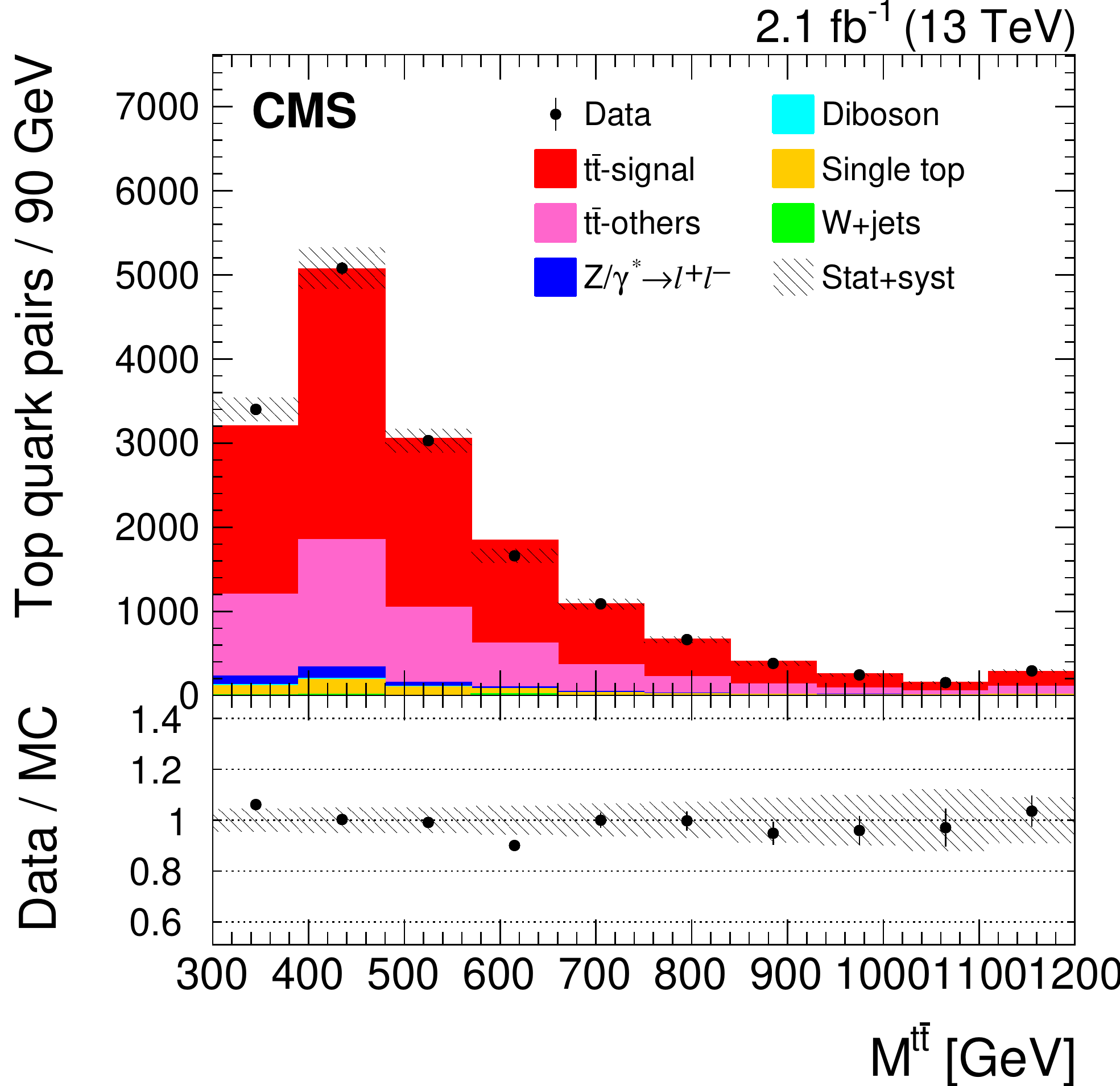}
\includegraphics[width=\cmsFigWidth]{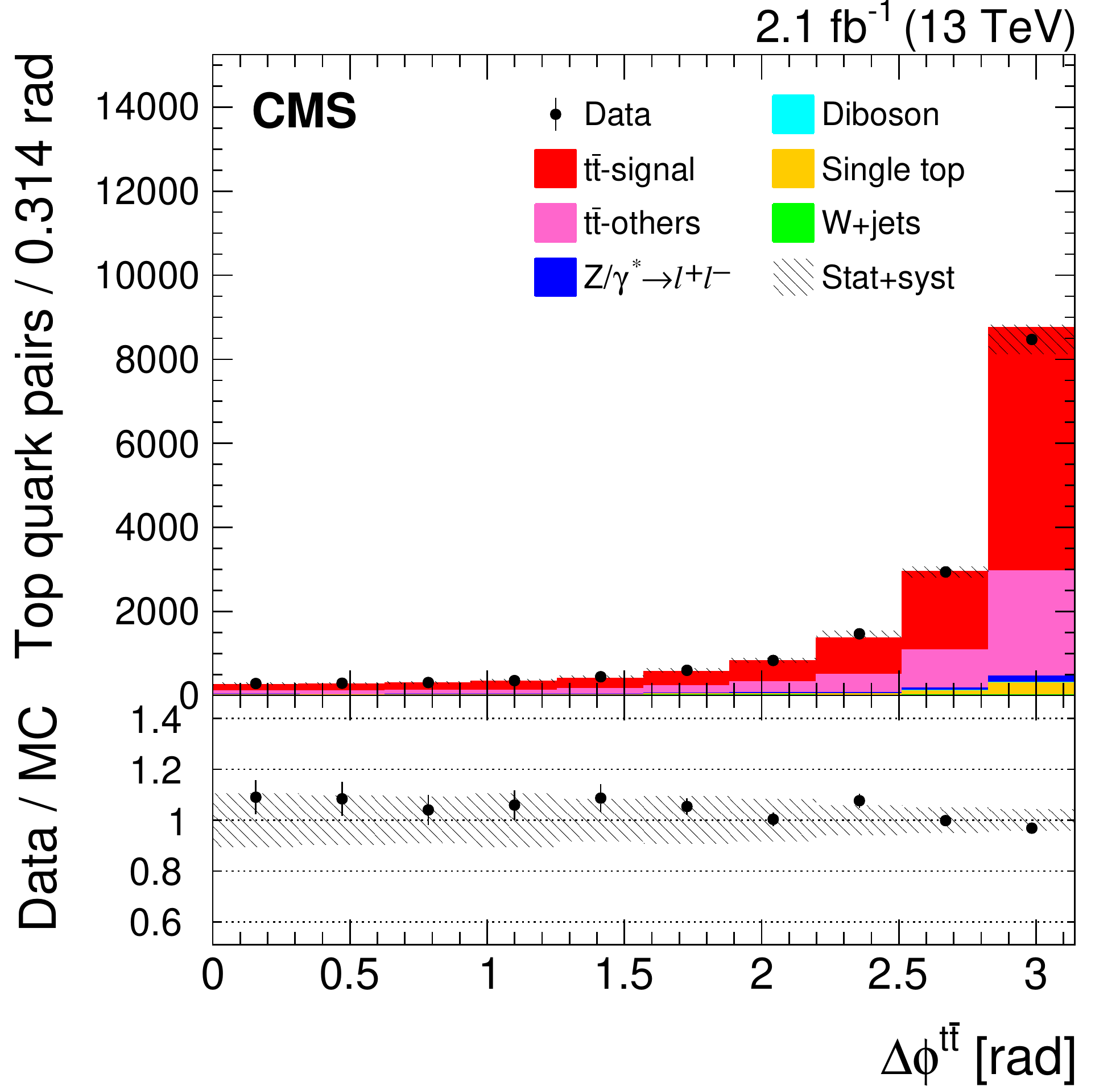}
\caption{Reconstructed $\pttt$ (upper left), $\ytt$ (upper right), $\mtt$ (lower left), and $\dphtt$ (lower right) distributions from data (points) and from MC simulation (shaded histograms). The signal definition for particle level is considered to distinguish \ttbar-signal and \ttbar-others. All corrections described in the text are applied to the simulation. The last bin includes the overflow events. The uncertainties shown by the vertical bars on the data points are statistical only while the hatched band shows the combined statistical and systematic uncertainties added in quadrature. The lower panels display the ratios of the data to the MC prediction.}
\label{fig:dist3}
\vspace{10pt}
\end{figure}

The kinematic variables of the leptons, \PQb~jets, top quarks, and \ttbar system are taken from the selected final solution.
Figure~\ref{fig:dist1} shows the distributions of the transverse momenta of leptons ($\ptlep$), jets ($\ptjet$), and top quarks ($\ptt$), and the rapidity of the top quarks ($\yt$). Figure~\ref{fig:dist3} displays the distributions of the transverse momentum ($\pttt$), rapidity ($\ytt$), and invariant mass ($\mtt$) of the ${\ttbar}$ system, and the azimuthal angle between the top quarks ($\dphtt$). In the upper panel of each figure, the data points are compared to the sum of the expected contributions obtained from MC simulated events reconstructed as the data. The lower panel shows the ratio of the data to the expectations. The measured $\ptlep$, $\ptjet$, and $\ptt$ distributions are softer than those predicted by the MC simulation, resulting in the negative slopes observed in the bottom panels.  However, in general, there is reasonable agreement between the data and simulation within the uncertainties, which are discussed in Sec.~\ref{sec:error}.

\section{Normalized differential cross sections}
\label{sec:DiffXsec}

The normalized differential \ttbar cross sections $ (1/\sigma)( \rd \sigma / \rd X ) $ are measured as a function of several different kinematic variables $X$. The variables include $\ptt$, $\pttt$, $\yt$, $\ytt$,  $\mtt$, and $\dphtt$, at both the particle and parton levels. In addition, the measurements are performed with $\ptlep$ and $\ptjet$ at the particle level. The measurements are compared to the predictions of ${\POWHEG}$+\PYTHIA{}8, {\aMCatNLO}+\PYTHIA{}8\FXFX, {\aMCatNLO}+\PYTHIA{}8\MLM, and ${\POWHEG}$+${\HERWIGpp}$.

The non-\ttbar backgrounds are estimated from simulation and subtracted from the data. For Drell--Yan processes the normalization of the simulation is determined from the data using the ``$R_\text{out/in}$'' method~\cite{Khachatryan:2010ez,Chatrchyan:2011nb,Chatrchyan:2012bra}. The non-\ttbar backgrounds are first subtracted from the measured distributions. The data distributions are slightly lower than those from the MC simulation. The \ttbar-others backgrounds are then removed as a proportion of the total \ttbar contribution by applying a single correction factor $k$ shown in Eq.~(\ref{eq:extract}), using Eq. (\ref{eq:extract2}):

\begin{equation}\label{eq:extract}
k=\frac{N^{\text{data}}-N^{\mathrm{MC}}_{\text{non-}\ttbar}}{N^{\mathrm{MC}}_{\ttbar\text{-sig}} + N^{\mathrm{MC}}_{ \ttbar\text{-others}}},
\end{equation}
\begin{equation}\label{eq:extract2}
\begin{aligned}
N^{\text{data}}_{\ttbar\text{-sig}} &= N^{\text{data}} - N^{\mathrm{MC}}_{\text{non-}\ttbar} - k N^{\mathrm{MC}}_{\ttbar\text{-others}}.
\end{aligned}
\end{equation}
Here, $N^{\mathrm{MC}}_{\mathrm{non}\textnormal{-}\ttbar}$ is the total estimate for the non-${\ttbar}$ background from the MC simulation, $N^{\mathrm{MC}}_{\ttbar\text{-sig}}$ is the total MC-predicted ${\ttbar}$ signal yield, and $N^{\mathrm{MC}}_{\ttbar\text{-others}}$ is the total MC prediction of the remaining \ttbar background.
The \ttbar signal yield, $N^{\text{data}}_{\ttbar\textnormal{-}\mathrm{sig}}$, is then extracted from the number of data events, $N^{\text{data}}$, separately in each bin of the kinematic distributions, as shown in Eq.~(\ref{eq:extract2}).

The bin widths of the distributions are chosen to control event migration between the bins at the reconstruction and generator level due to detector resolutions.
We define the purity (stability) as the number of events generated and correctly reconstructed in a certain bin, divided by the total number of events in the reconstruction-level (generator-level) bin.  The bin widths are chosen to give both a purity and a stability of about 50\%.

Detector resolution and reconstruction efficiency effects are corrected using an unfolding procedure. The method relies on a response matrix that maps the expected relation between the true and reconstructed variables taken from the ${\POWHEG}$+\PYTHIA{}8 simulation. The D'Agostini method~\cite{DAgostini:1994fjx} is employed to perform the unfolding. The effective regularization strength of the iterative D'Agostini unfolding is controlled by the number of iterations. A small number of iterations can bias the measurement towards the simulated prediction, while with a large number of iterations the result converges to that of a matrix inversion. The number of iterations is optimized for each distribution, using simulation to find the minimum number of iterations that reduces the bias to a negligible level. This optimization is performed with the multiplication of the response matrix and does not require any regularization. A detailed description of the method can be found in Ref.~\cite{Khachatryan:2016mnb}.

\section{Systematic uncertainties}
\label{sec:error}
Several sources of systematic uncertainties are studied. The normalized differential cross sections are remeasured with respect to each source of systematic uncertainty individually, and the differences from the nominal values in each bin are taken as the corresponding systematic uncertainty. The overall systematic uncertainties are then obtained as the quadratic sum of the individual components.

The pileup distribution used in the simulation is varied by shifting the assumed total inelastic pp cross section by ${\pm}5\%$, in order to determine the associated systematic uncertainty.  The systematic uncertainties in the lepton trigger, identification, and isolation efficiencies are determined by varying the measured scale factors by their total uncertainties. Uncertainties coming from the jet in the jet energy scale (JES) and jet energy resolution (JER) are determined on a per-jet basis by shifting the energies of the jets~\cite{Chatrchyan:2011ds} within their measured energy scale and resolution uncertainties. The b tagging uncertainty is estimated by varying its efficiency uncertainty.

The uncertainty in the non-\ttbar background normalization is estimated using a 15--30\% variation in the background yields, which is based on a previous CMS measurement of the \ttbar cross section~\cite{Khachatryan:2016kzg}. The uncertainty in the shape of the \ttbar-others contribution is obtained by reweighting the \pt distribution of the top quark for the \ttbar-others events to match the data and comparing with the unweighted contribution. For the theoretical uncertainties, we investigate the effect of the choice of PDFs, factorization and renormalization scales ($\mu_\mathrm{F}$ and $\mu_\mathrm{R}$), variation of the top quark mass, top quark \pt, and hadronization and generator modeling.

The PDF uncertainty is estimated using the uncertainties in the NNPDF30\_NLO\_as\_0118 set with the strong coupling strength $\alpha_\mathrm{s} = 0.118 $~\cite{Ball:2014uwa}. We measure 100 individual uncertainties and take the root-mean-square as the PDF uncertainty, following the PDF4LHC recommendation~\cite{Butterworth:2015oua}. In addition, we consider the PDF sets with $\alpha_\mathrm{s} = 0.117$ and 0.119. The MC generator modeling uncertainties are estimated by taking the difference between the results based on the \POWHEG and \aMCatNLO generators.

The uncertainty from the choice of $\mu_\mathrm{F}$ and $\mu_\mathrm{R}$ is estimated by varying the scales by a factor of two up and down in \POWHEG independently for the ME and PS steps. For the ME calculation, all possible combinations are considered independently, excluding the most extreme cases of ($\mu_\mathrm{F}$, $\mu_\mathrm{R}$) = (0.5, 2) and (2, 0.5)~\cite{Cacciari:2003fi,Catani:2003zt}. The scale uncertainty in the PS modeling is assessed using dedicated MC samples with the scales varied up and down together. The uncertainties in the factorization and renormalization scales in the ME and PS calculations are taken as the envelope of the differences with respect to the nominal parameter choice.

We evaluate the top quark mass uncertainty by taking the maximum deviation between the nominal MC sample with a top quark mass of 172.5\GeV and samples with masses of 171.5 and 173.5\GeV.
The \ttbar signal cross sections are not corrected for the mismodeling of the top quark \pt distribution in simulation. Instead, a systematic uncertainty from this mismodeling is obtained by comparing the nominal results to the results obtained from a response matrix using \ttbar-signal in which the top quark \pt distribution is reweighted to match the data. The uncertainty from hadronization and PS modeling is estimated by comparing the results obtained from \POWHEG samples interfaced with \PYTHIA{}8 and with \HERWIGpp.

Table~\ref{tab:sys} lists typical values for the statistical and systematic uncertainties in the measured normalized \ttbar differential cross sections. The table gives the uncertainty sources and corresponding range of the median uncertainty of each distribution, at both the particle and parton levels. The hadronization is the dominant systematic uncertainty source for $\ptt$ ($4.9\%$ at particle and $7.1\%$ at parton level) and $\mtt$ ($5.9\%$ at particle and $7.4\%$ at parton level), and the MC generator modeling is dominant for $\yt$ ($2.3\%$ at particle and $2.2\%$ at parton level), $\pttt$ ($6.1\%$ at particle and $3.9\%$ at parton level), $\ytt$ ($1.2\%$ at particle and $1.6\%$ at parton level), and $\dphtt$ ($9.2\%$ at particle and $7.3\%$ at parton level). In general, the MC generator modeling and hadronization are the dominant systematic uncertainty sources for both the particle- and parton-level measurements.

\begin{table}[htb]
\centering
\topcaption{Statistical and systematic uncertainties in the normalized \ttbar differential cross sections at particle and parton levels. The uncertainty sources and the corresponding range of the median uncertainty of each distribution are shown in percent.}
\label{tab:sys}
\renewcommand{\arraystretch}{1.1}
\begin{tabular}{l |c | c}
Uncertainty source & Particle level [\%] & Parton level [\%] \\ \hline
Statistical &  0.24 -- 0.59  & 0.36 -- 0.63 \\  \hline
Pileup modeling&  0.02 -- 0.48 &  0.07 -- 0.49 \\
Trigger efficiency &   0.03 -- 0.67  & 0.06 -- 0.82 \\
Lepton efficiency &  0.06 -- 0.94 & 0.07 -- 0.90 \\
JES &   0.14 -- 2.04   & 0.29 -- 1.44 \\
JER &   0.04 -- 0.85  & 0.29 -- 0.65 \\
\PQb~jet tagging &  0.12 -- 1.19 &  0.26 -- 1.16 \\
Background &  0.13 -- 2.14 & 0.09 -- 1.28 \\
PDFs &   0.15 -- 0.96 &  0.17 -- 0.97 \\
MC generator &  0.66 -- 9.24 & 1.61 -- 7.32 \\
Fact./renorm. &  0.10 -- 4.15  & 0.17 -- 4.15 \\
Top quark mass &  0.49 -- 1.89  & 0.68 -- 3.05 \\
Top quark $\pt$ & 0.02 -- 1.74 & 0.02 -- 0.69 \\
Hadronization --- PS modeling &   0.70 -- 5.85 & 0.41 -- 7.44 \\ \hline
Total systematic uncertainty   &  1.7 -- 15\y &  3.1 -- 13\y \\
\end{tabular}
\end{table}

\section{Results}

\begin{figure}
\centering
\includegraphics[width=\cmsFigWidth]{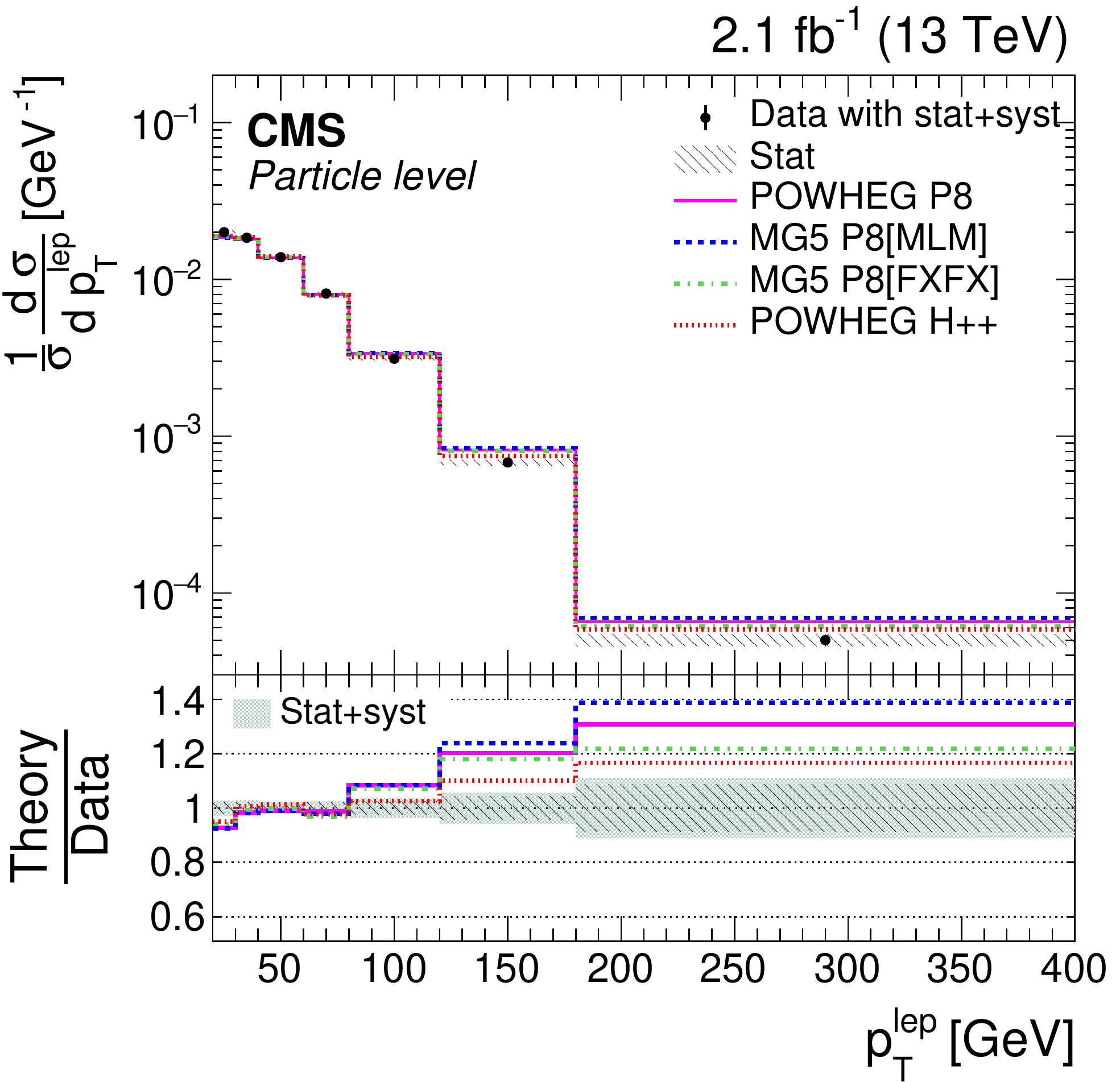}
\includegraphics[width=\cmsFigWidth]{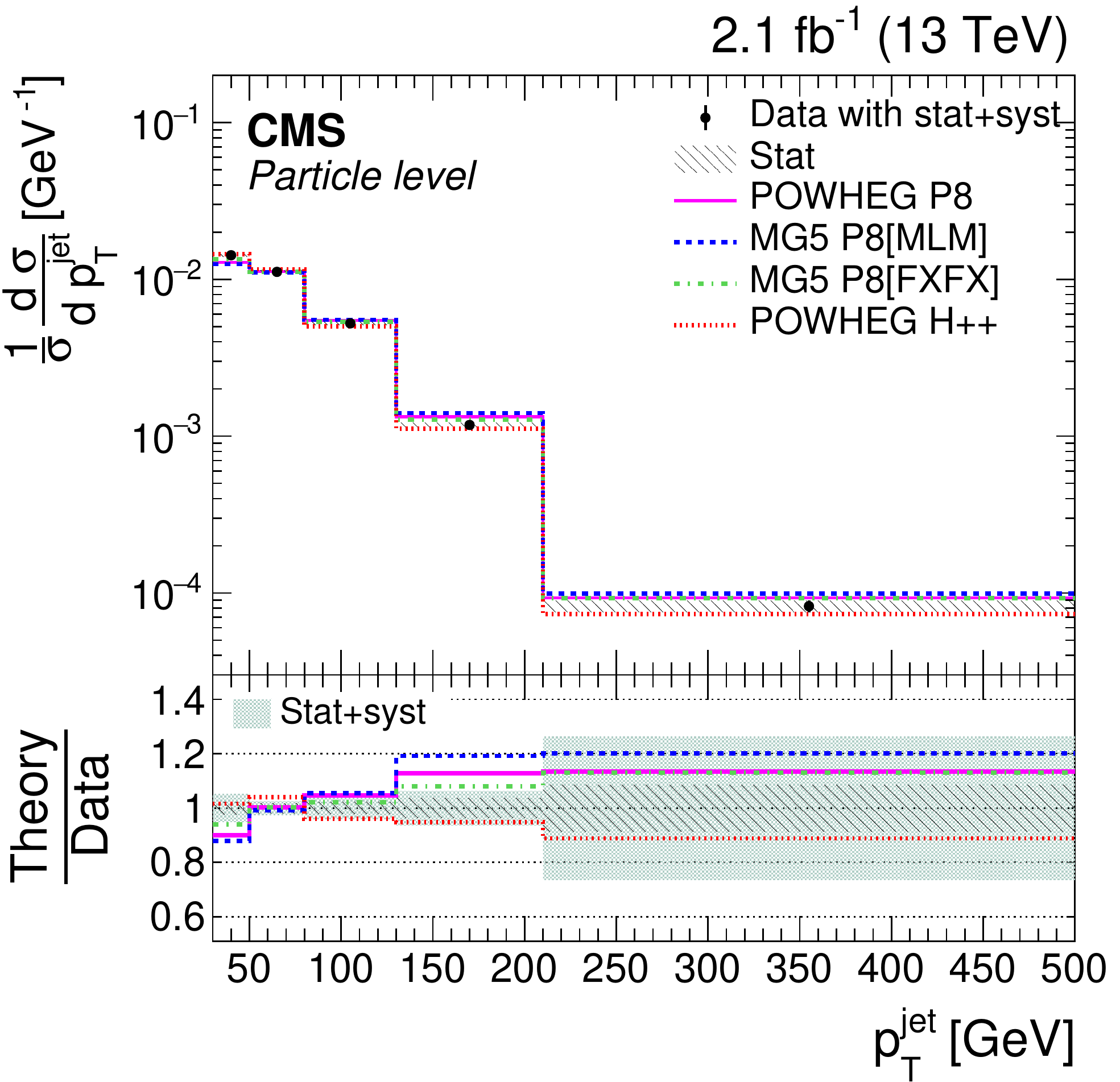}
\\ \vspace{0.2cm}
\includegraphics[width=\cmsFigWidth]{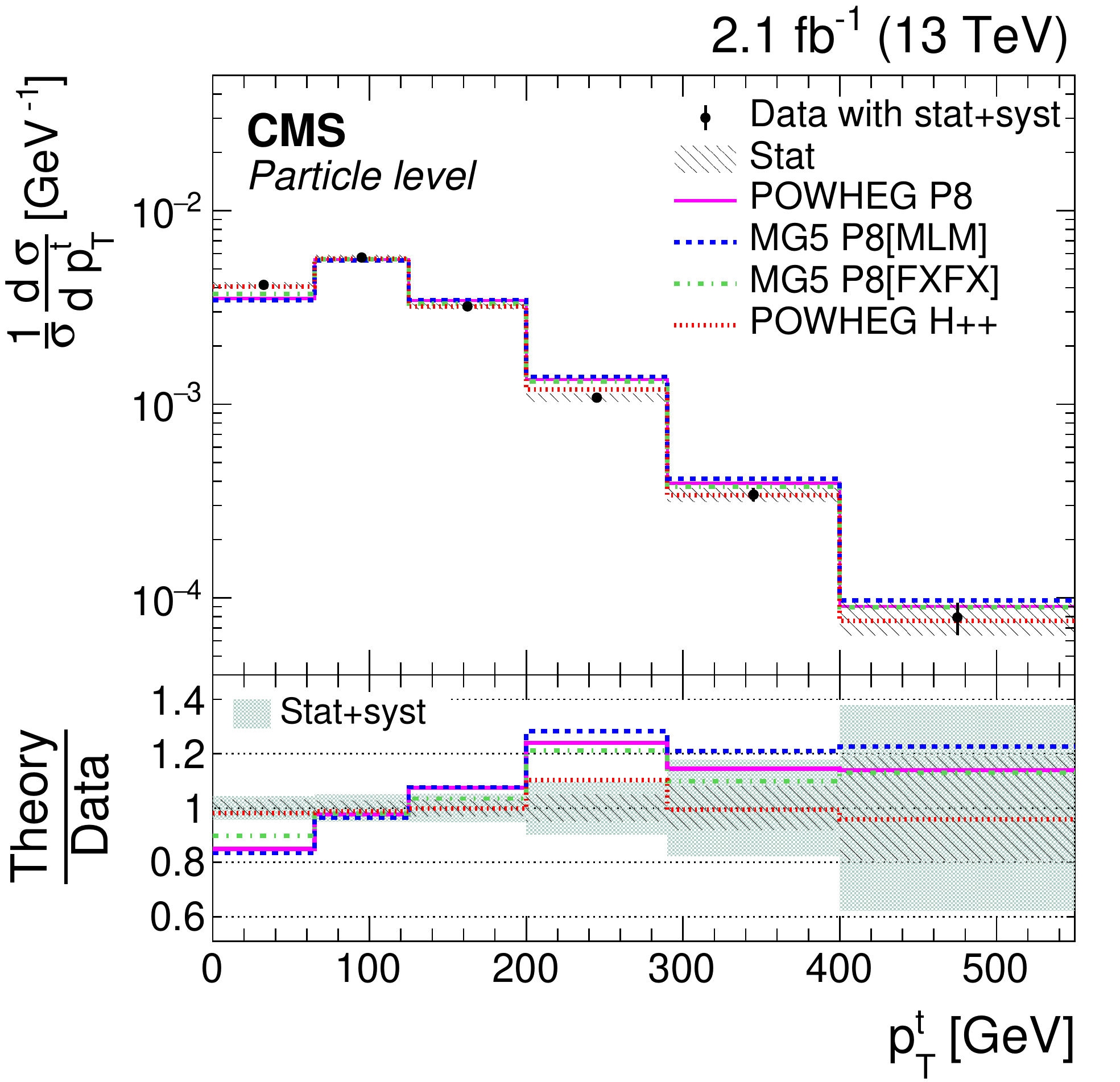}
\includegraphics[width=\cmsFigWidth]{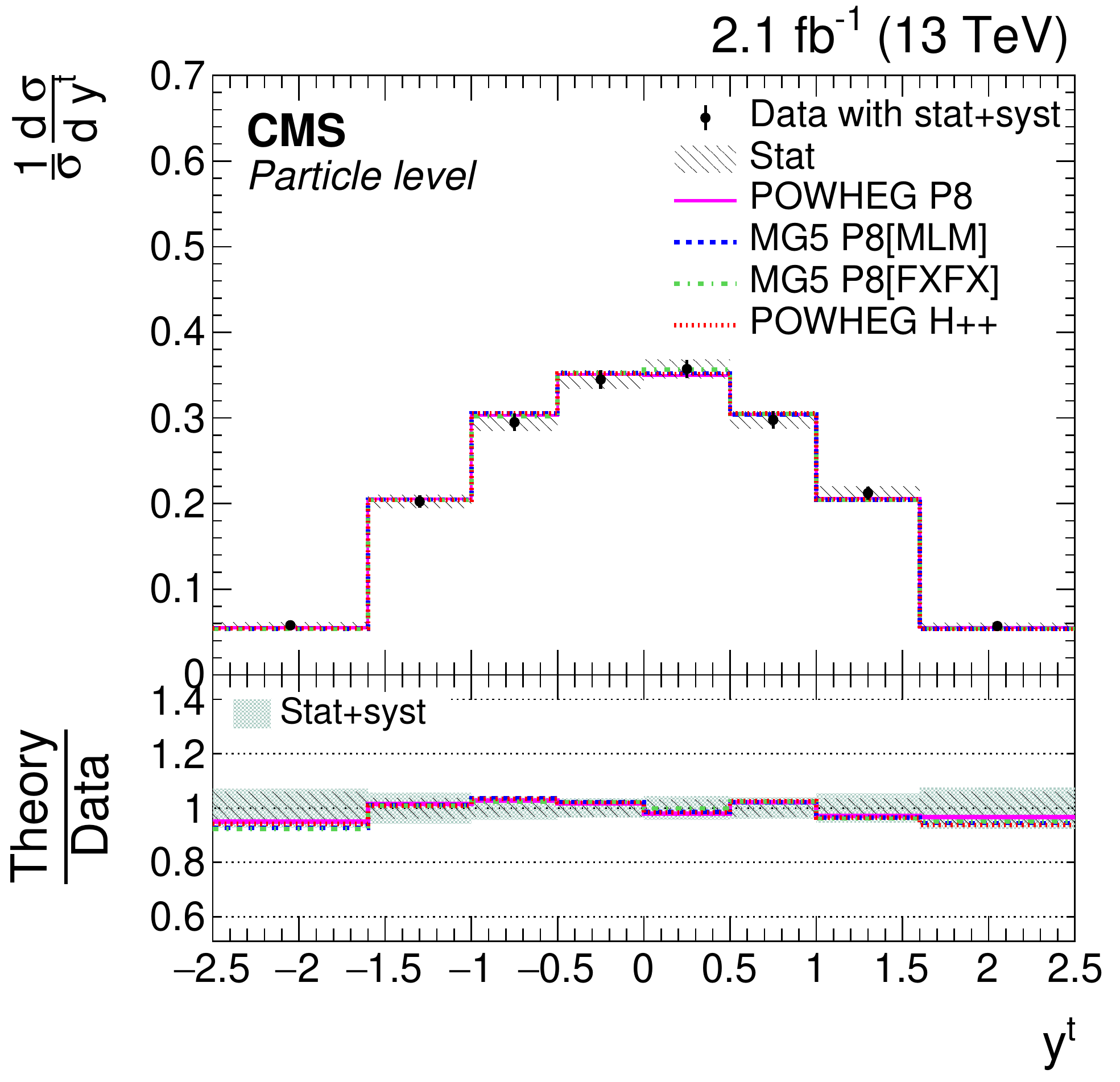}
\caption{Normalized differential \ttbar cross sections as a function of lepton (upper left), jet (upper right), and top quark \pt (lower left) and top quark rapidity (lower right), measured at the particle level in the visible phase space and combining the distributions for top quarks and antiquarks. The measured data are compared to different standard model predictions from ${\POWHEG}$+\PYTHIA{}8 (\cmsSF{POWHEG~P8}), {\aMCatNLO}+\PYTHIA{}8\MLM  (\cmsSF{MG5 P8[MLM]}), {\aMCatNLO}+\PYTHIA{}8\FXFX (\cmsSF{MG5 P8[FXFX]}), and ${\POWHEG}$+${\HERWIGpp}$ (\cmsSF{POWHEG H++}). The vertical bars on the data points indicate the total (combined statistical and systematic) uncertainties while the hatched band shows the statistical uncertainty. The lower panel gives the ratio of the theoretical predictions to the data. The light-shaded band displays the combined statistical and systematic uncertainties added in quadrature.}
\label{fig:particle_level_1}
\vspace{10pt}
\end{figure}

\begin{figure}
\centering
\includegraphics[width=\cmsFigWidth]{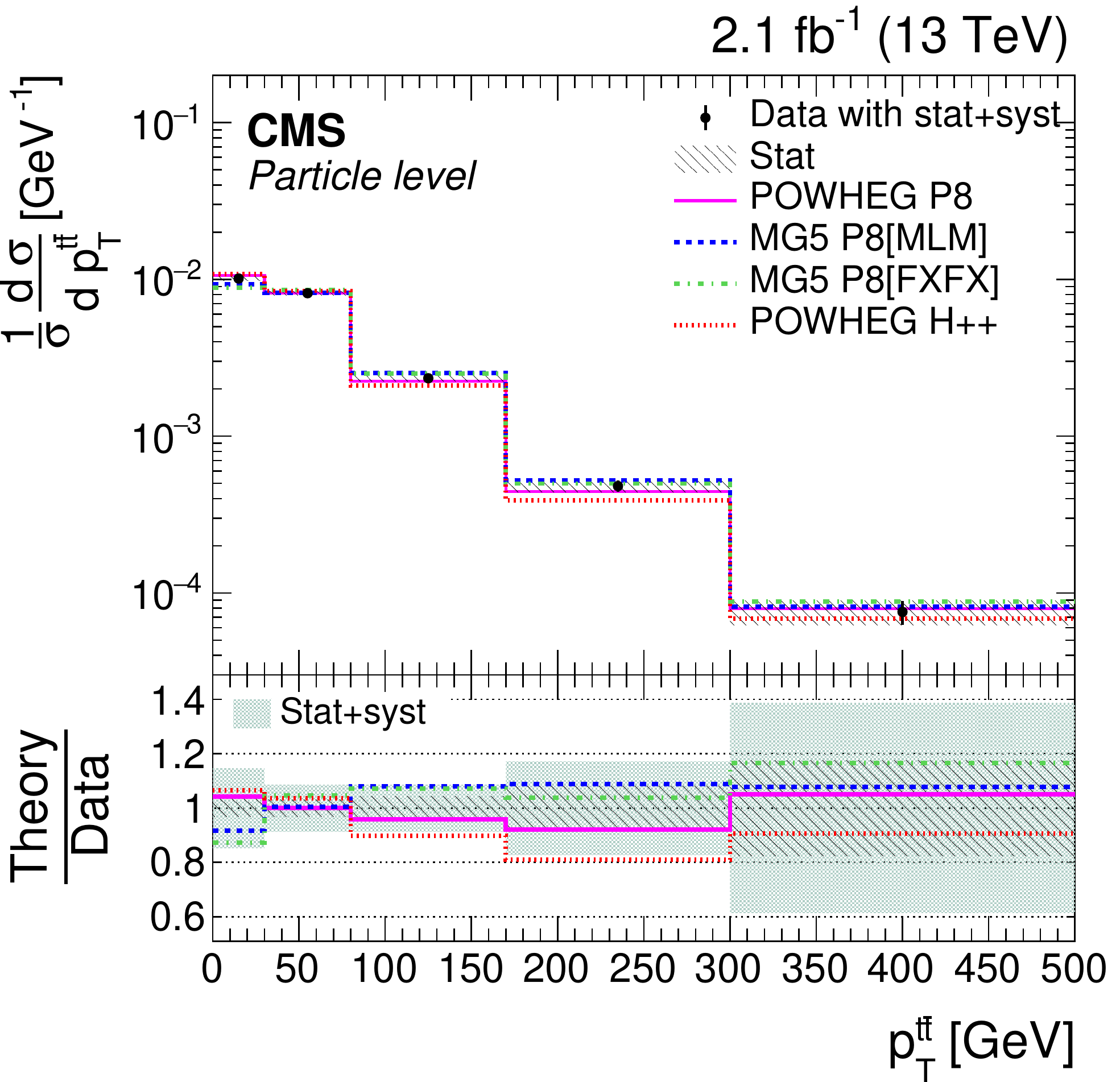}
\includegraphics[width=\cmsFigWidth]{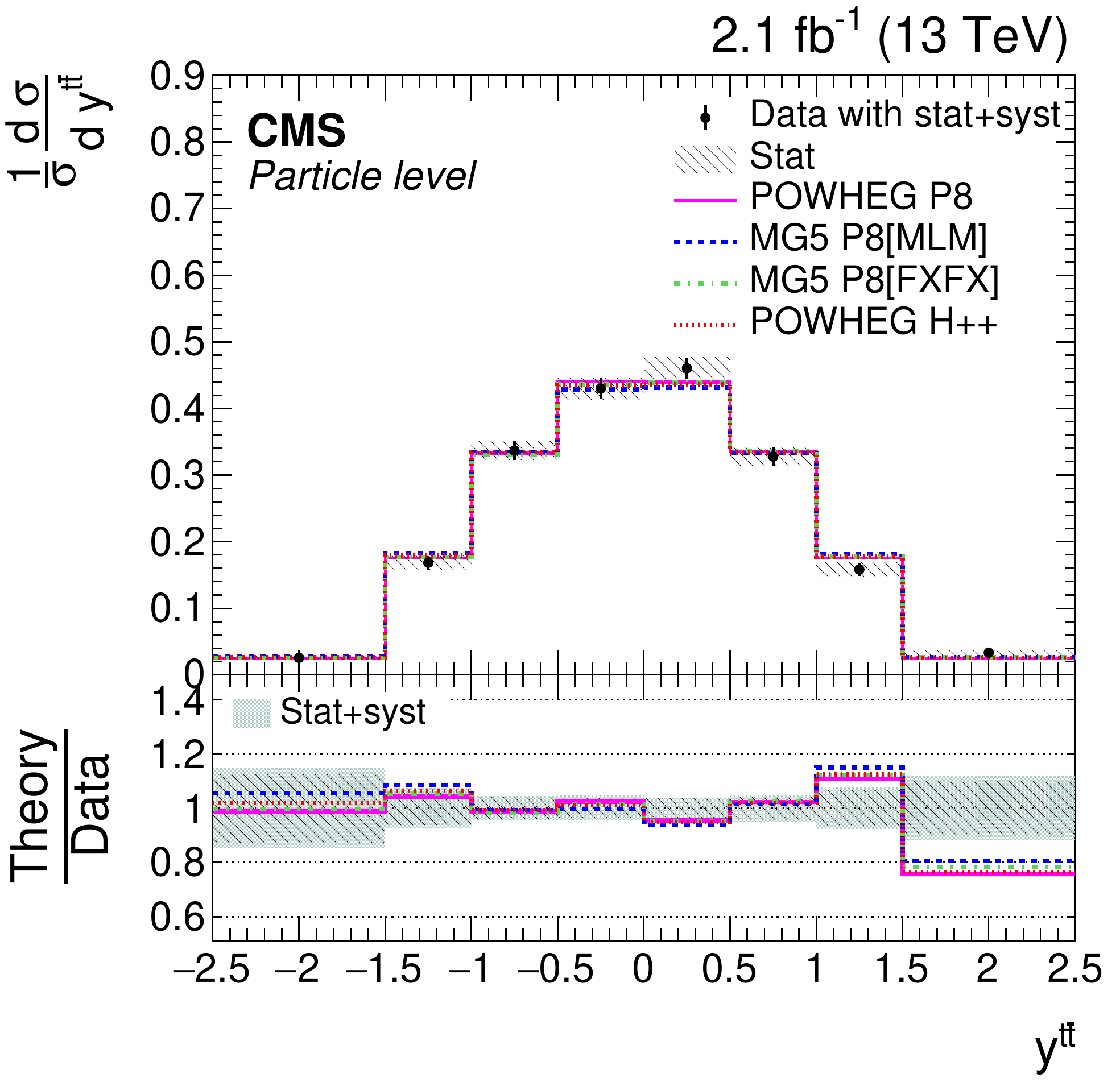}
\\ \vspace{0.2cm}
\includegraphics[width=\cmsFigWidth]{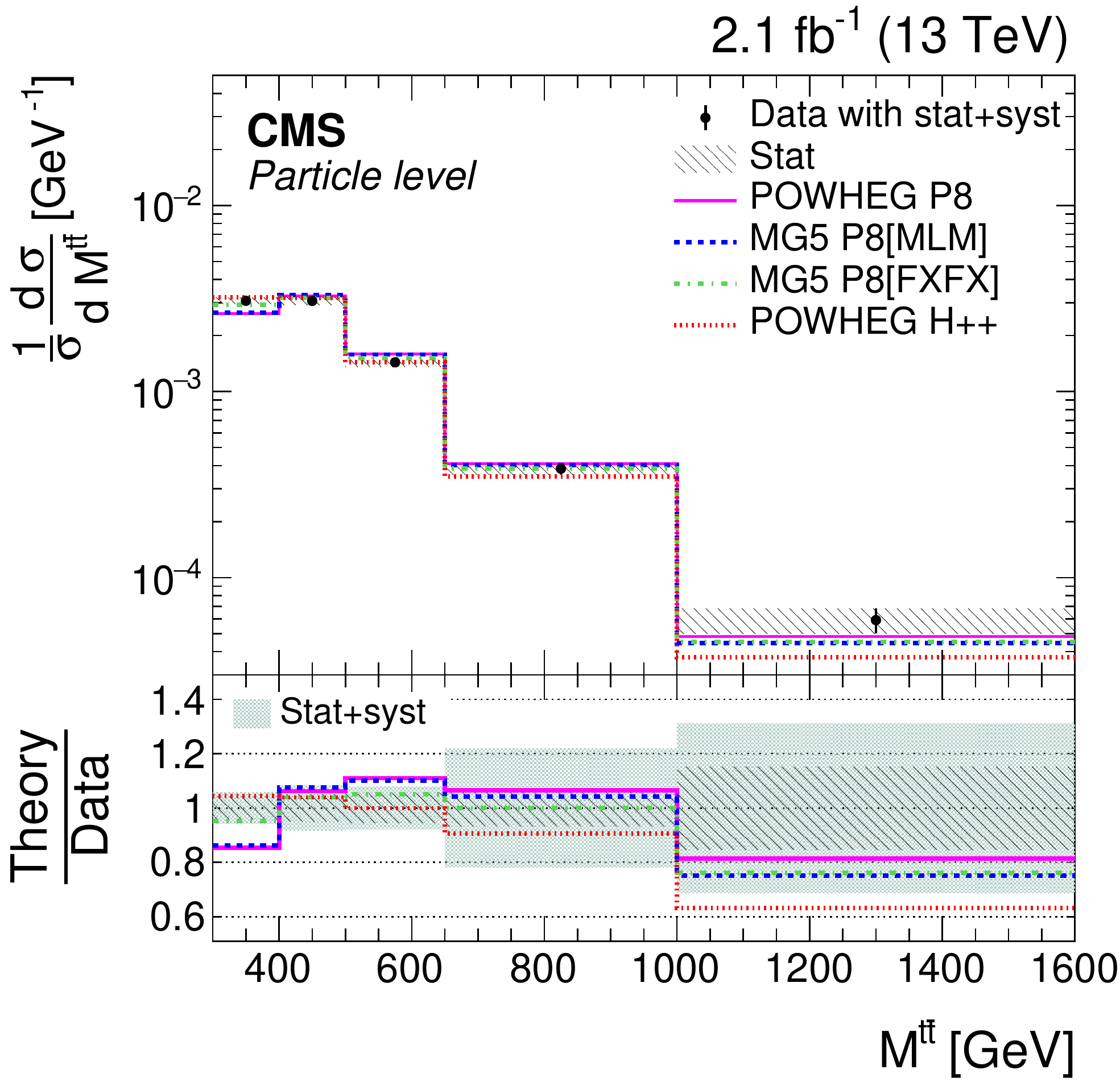}
\includegraphics[width=\cmsFigWidth]{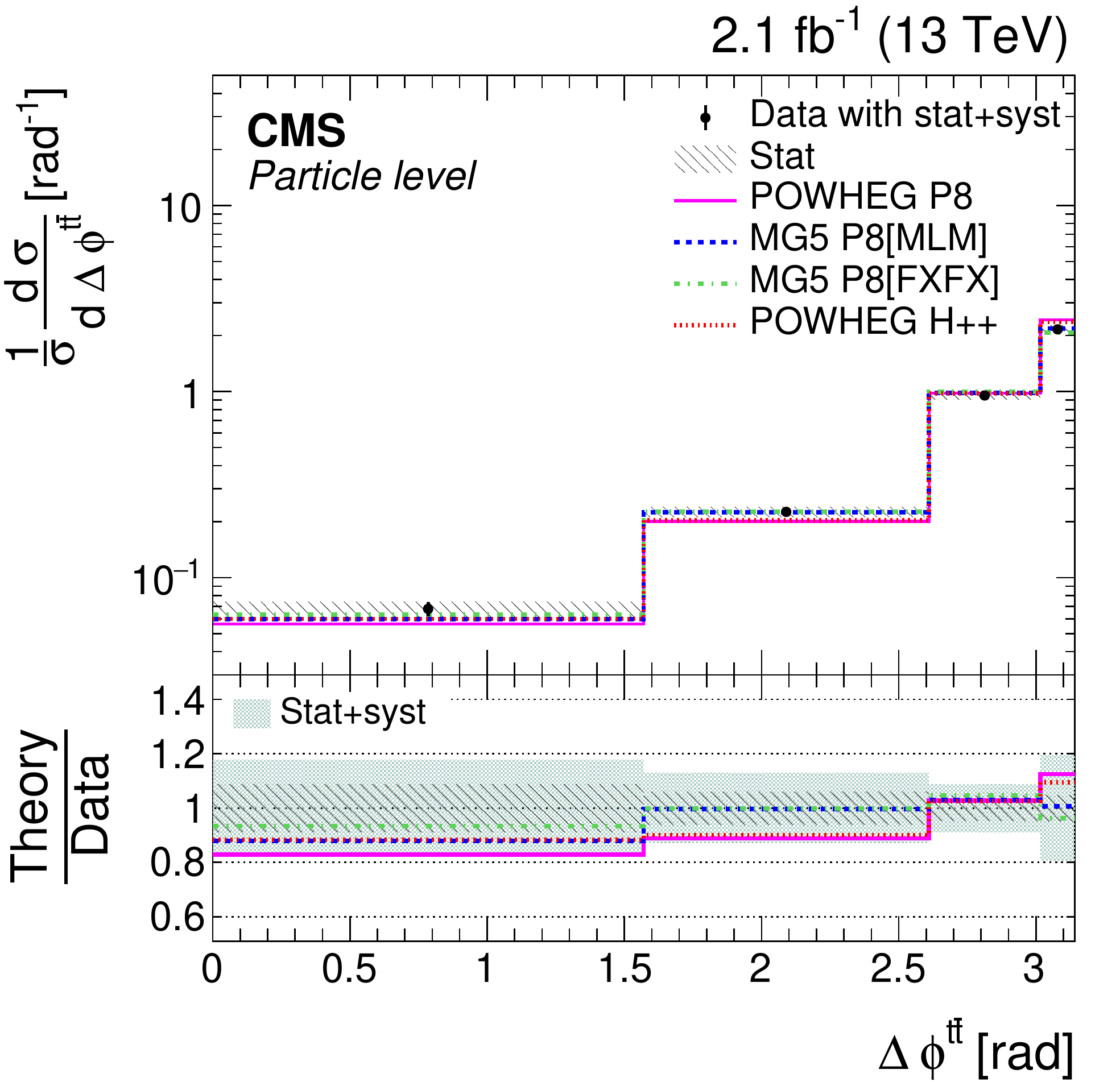}
\caption{Normalized differential \ttbar cross sections as a function of $\pttt$ (upper left), $\ytt$ (upper right),  $\mtt$ (lower left), and $\dphtt$ (lower right), measured at the particle level in the visible phase space. The measured data are compared to different standard model predictions from ${\POWHEG}$+\PYTHIA{}8 (\cmsSF{POWHEG~P8}), {\aMCatNLO}+\PYTHIA{}8\MLM  (\cmsSF{MG5 P8[MLM]}), {\aMCatNLO}+\PYTHIA{}8\FXFX (\cmsSF{MG5 P8[FXFX]}), and ${\POWHEG}$+${\HERWIGpp}$ (\cmsSF{POWHEG H++}). The vertical bars on the data points indicate the total (combined statistical and systematic) uncertainties while the hatched band shows the statistical uncertainty. The lower panel gives the ratio of the theoretical predictions to the data. The light-shaded band displays the combined statistical and systematic uncertainties added in quadrature.}
\label{fig:particle_level_2}
\vspace{10pt}
\end{figure}

\begin{figure}
\centering
\includegraphics[width=\cmsFigWidth]{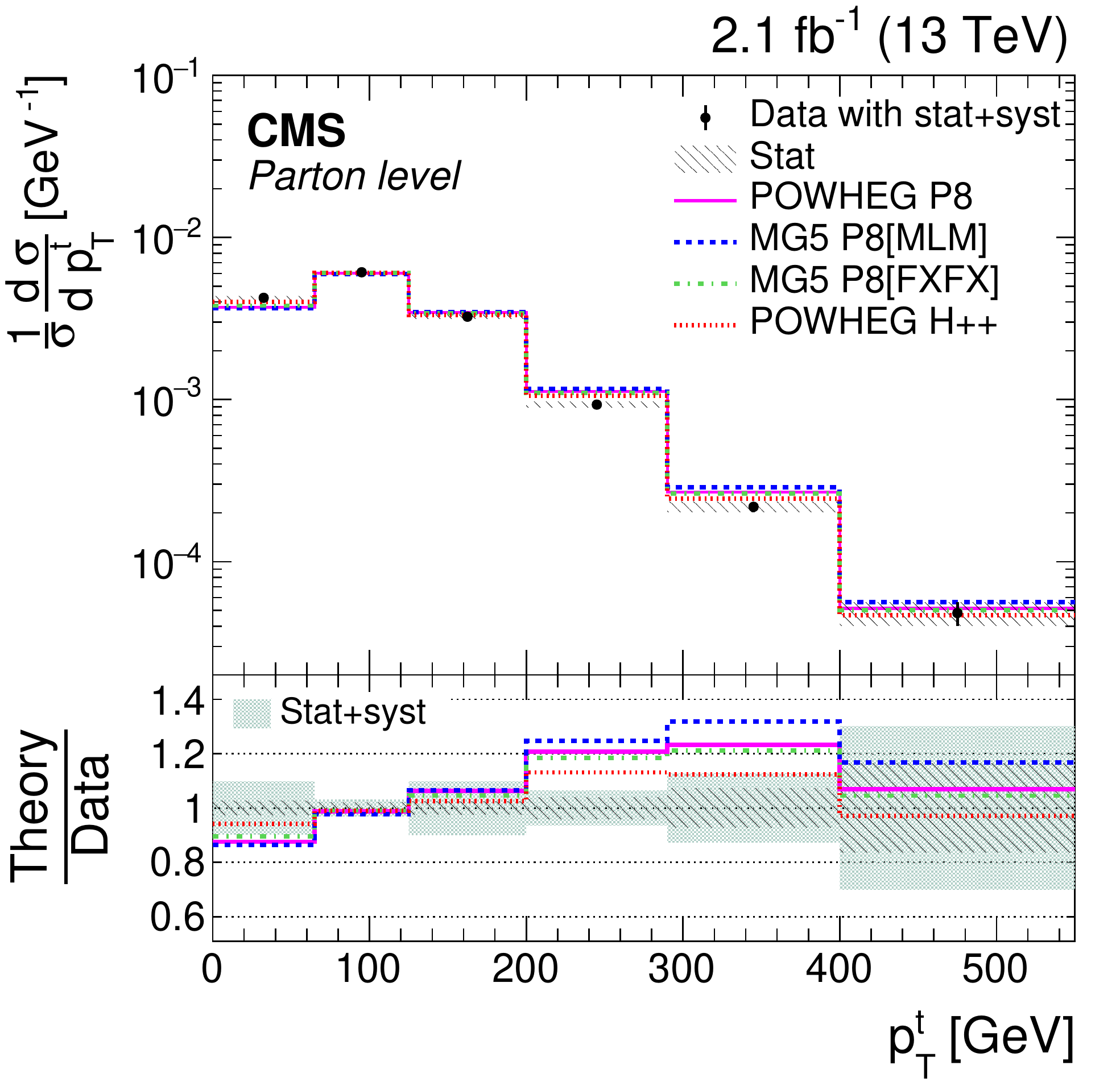}
\includegraphics[width=\cmsFigWidth]{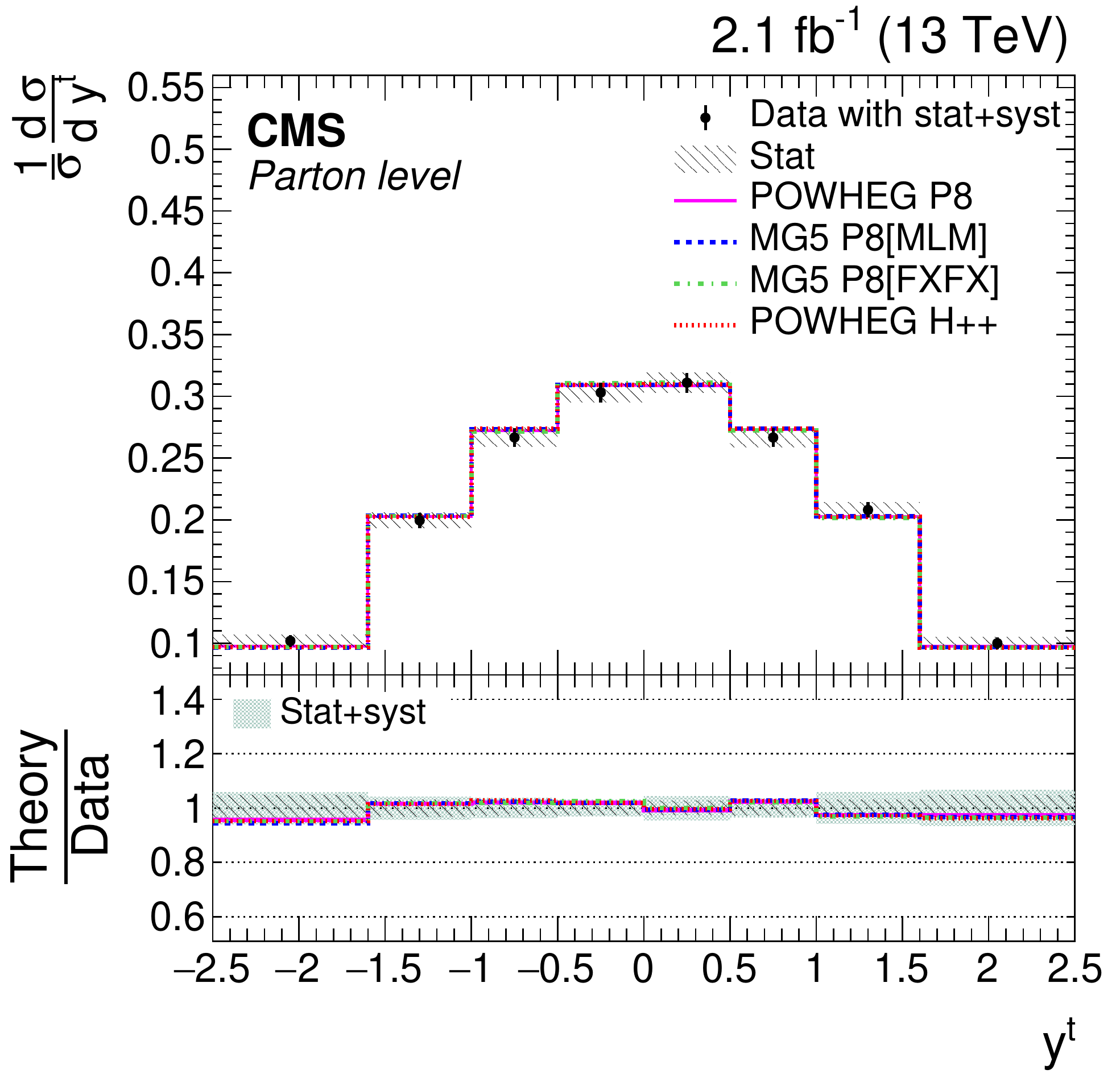}
\caption{Normalized differential \ttbar cross sections as a function of top quark \pt (left) and top quark rapidity (right), measured at the parton level in the full phase space and combining the distributions for top quarks and antiquarks. The measured data are compared to different standard model predictions from ${\POWHEG}$+\PYTHIA{}8 (\cmsSF{POWHEG~P8}), {\aMCatNLO}+\PYTHIA{}8\MLM  (\cmsSF{MG5 P8[MLM]}), {\aMCatNLO}+\PYTHIA{}8\FXFX (\cmsSF{MG5 P8[FXFX]}), and ${\POWHEG}$+${\HERWIGpp}$ (\cmsSF{POWHEG H++}). The vertical bars on the data points indicate the total (combined statistical and systematic) uncertainties while the hatched band shows the statistical uncertainty. The lower panel gives the ratio of the theoretical predictions to the data. The light-shaded band displays the combined statistical and systematic uncertainties added in quadrature.}
\label{fig:parton_level_1}
\vspace{10pt}
\end{figure}

\begin{figure}
\centering
\includegraphics[width=\cmsFigWidth]{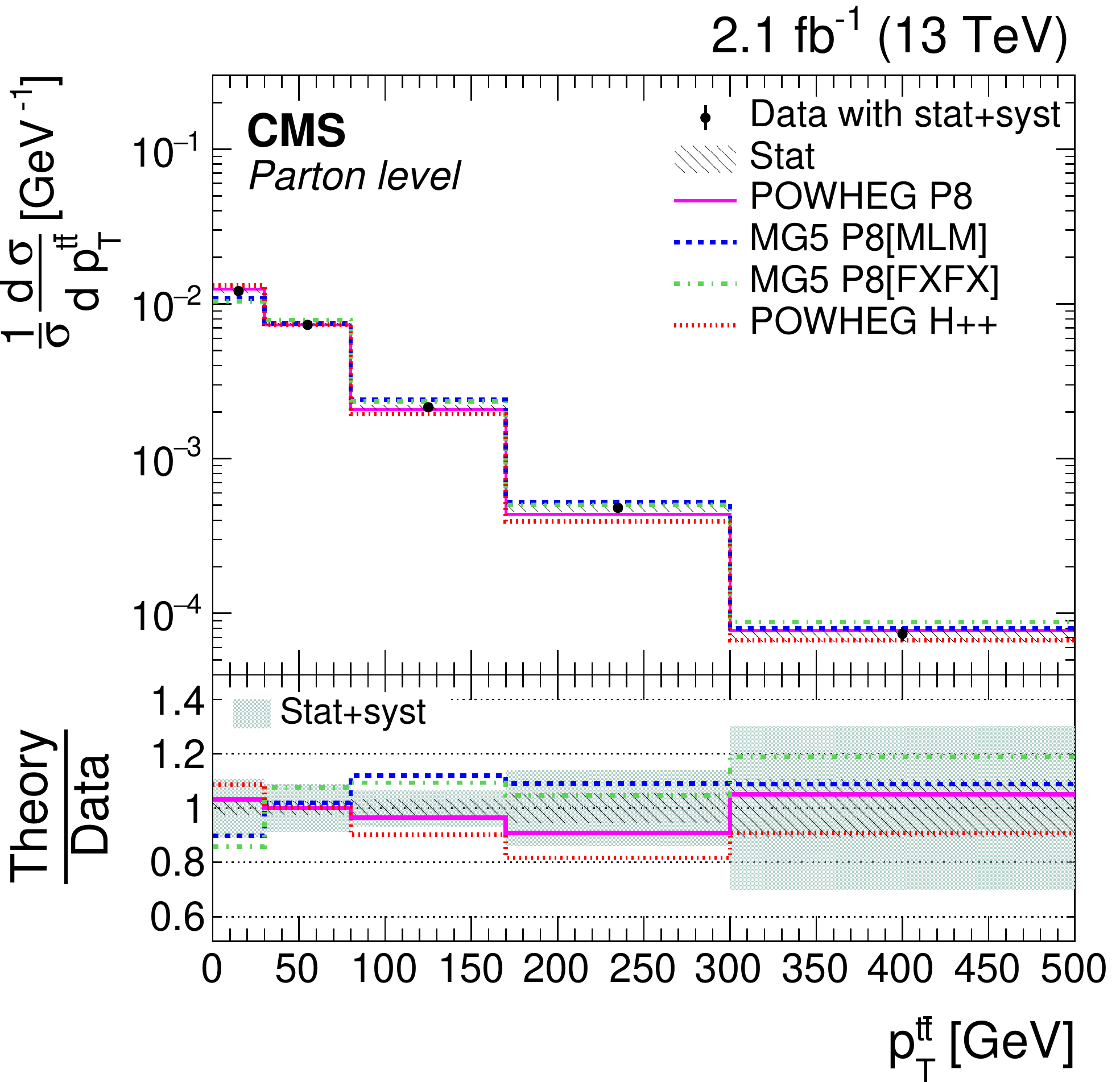}
\includegraphics[width=\cmsFigWidth]{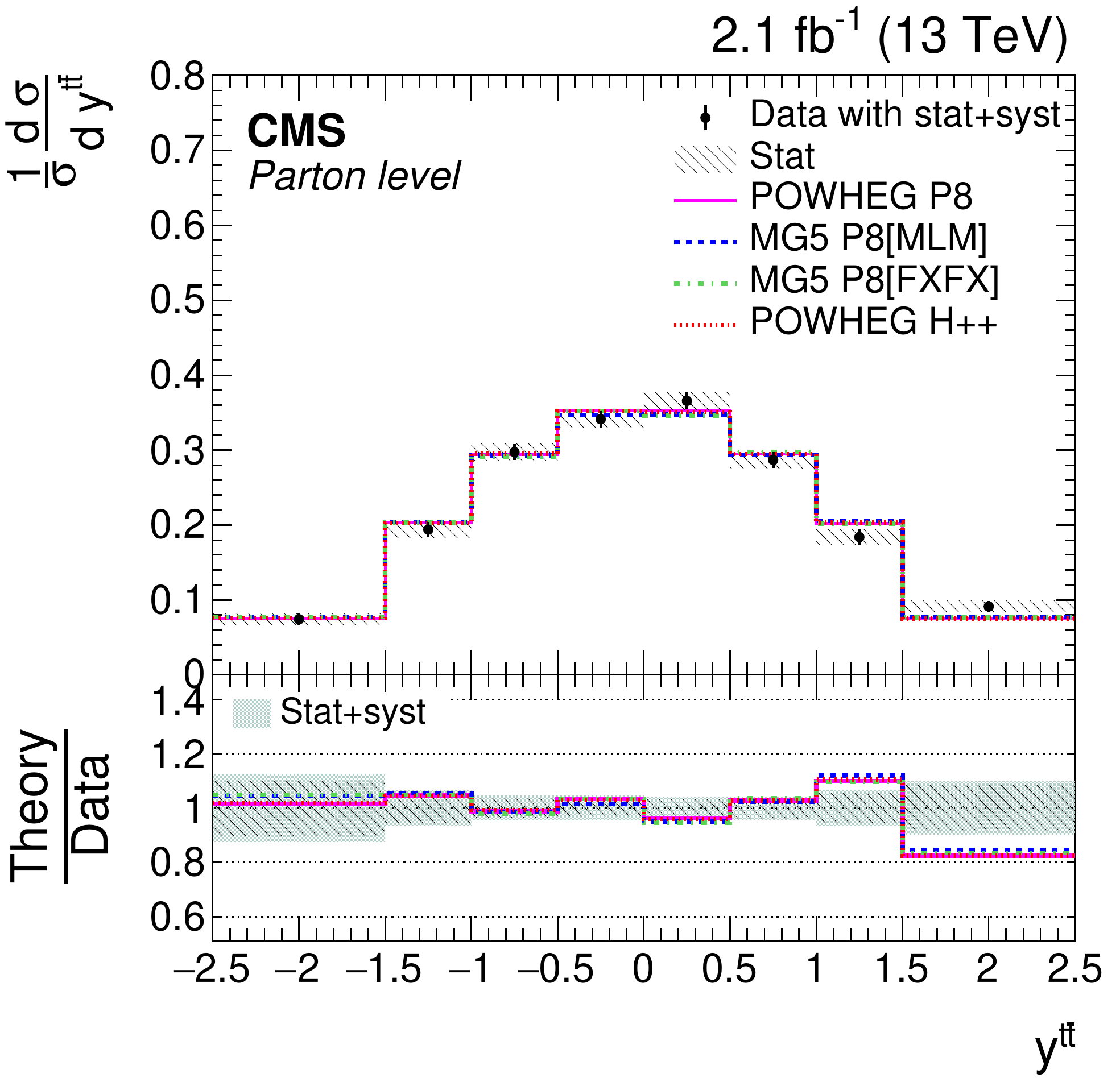}
\\ \vspace{0.2cm}
\includegraphics[width=\cmsFigWidth]{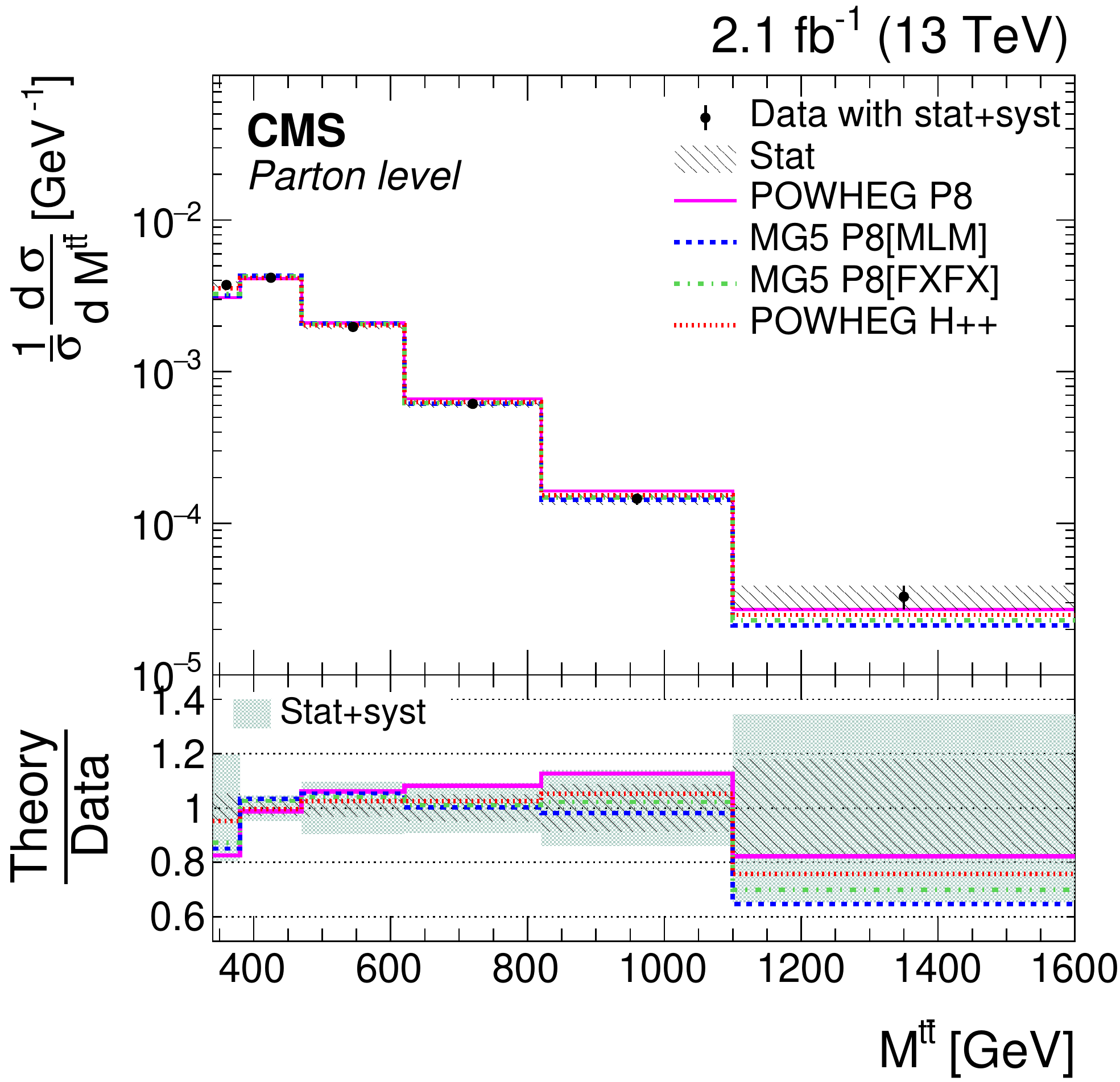}
\includegraphics[width=\cmsFigWidth]{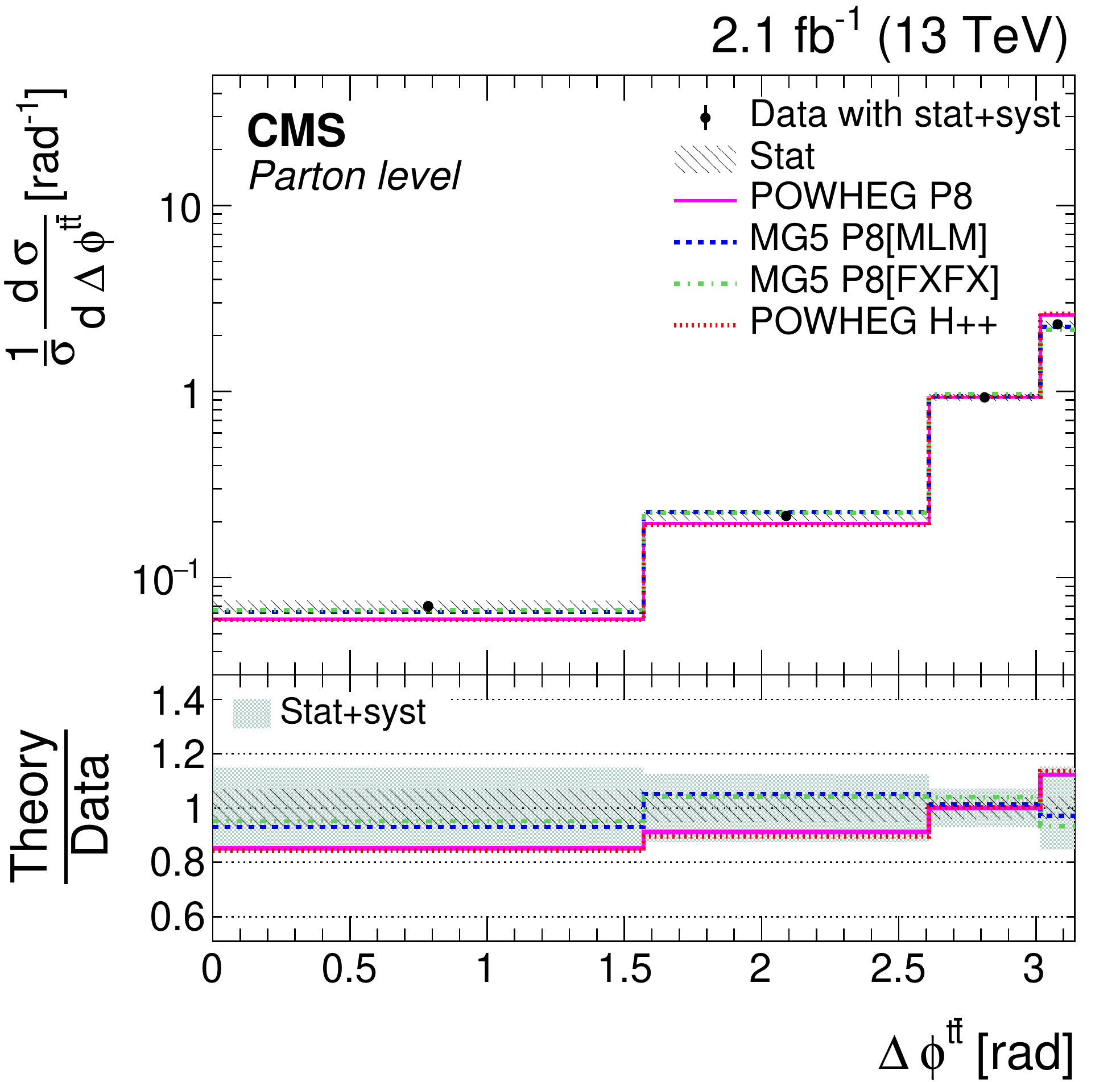}
\caption{Normalized differential \ttbar cross sections as a function of $\pttt$ (upper left), $\ytt$ (upper right),  $\mtt$ (lower left), and $\dphtt$ (lower right), measured at the parton level in the full phase space. The measured data are compared to different standard model predictions from ${\POWHEG}$+\PYTHIA{}8 (\cmsSF{POWHEG~P8}), {\aMCatNLO}+\PYTHIA{}8\MLM  (\cmsSF{MG5 P8[MLM]}), {\aMCatNLO}+\PYTHIA{}8\FXFX (\cmsSF{MG5 P8[FXFX]}), and ${\POWHEG}$+${\HERWIGpp}$ (\cmsSF{POWHEG H++}).  The vertical bars on the data points indicate the total (combined statistical and systematic) uncertainties while the hatched band shows the statistical uncertainty. The lower panel gives the ratio of the theoretical predictions to the data. The light-shaded band displays the combined statistical and systematic uncertainties added in quadrature.}
\label{fig:parton_level_2}
\vspace{10pt}
\end{figure}

{\tolerance=400
The normalized differential ${\ttbar}$ cross sections are measured by subtracting the background contribution, correcting for detector effects and acceptance, and dividing the resultant number of ${\ttbar}$ signal events by the total inclusive ${\ttbar}$ cross section.  Figs.~\ref{fig:particle_level_1} and \ref{fig:particle_level_2} show the normalized differential ${\ttbar}$ cross sections as a function of $\ptlep$, $\ptjet$, $\ptt$,  $\yt$, $\pttt$, $\ytt$,  $\mtt$, and $\dphtt$ at the particle level in the visible phase space. Parton-level results are also independently extrapolated to the full phase space using the ${\POWHEG}$+\PYTHIA{}8 \ttbar simulation.  Figures~\ref{fig:parton_level_1} and~\ref{fig:parton_level_2} show the normalized differential ${\ttbar}$ cross sections as a function of $\ptt$,  $\yt$, $\pttt$, $\ytt$,  $\mtt$, and $\dphtt$ at parton level in the full phase space.  The measured data are compared to different standard model predictions from ${\POWHEG}$+\PYTHIA{}8, {\aMCatNLO}+\PYTHIA{}8\FXFX, {\aMCatNLO}+\PYTHIA{}8\MLM, and ${\POWHEG}$+${\HERWIGpp}$ in the figures. The values of the measured normalized differential ${\ttbar}$ cross sections at the parton and particle levels with their statistical and systematic uncertainties are listed in Appendices A and B.
\par}

The compatibility between the measurements and the predictions is quantified by means of a $\chi^2$ test performed with the full covariance matrix from the unfolding procedure, including the systematic uncertainties. Tables~\ref{tab:chi2_gen1} and~\ref{tab:chi2_gen2} report the values obtained for the $\chi^{2}$ with the numbers of degrees of freedom (dof) and the corresponding p-values~\cite{Gross:2010qma}. The lepton, jet, and top quark \pt spectra in data tend to be softer than the MC predictions for the high-\pt region.  A similar trend was also observed at $\sqrt{s} = 8\TeV$ by both the ATLAS and CMS experiments~\cite{Khachatryan:2015oqa,Aad:2015mbv}. The ${\POWHEG}$+\PYTHIA{}8 generator better describes the $\pttt$,  $\yt$, and  $\ytt$ distributions at the particle and parton levels, while ${\POWHEG}$+\HERWIGpp is found to be in good agreement for the $\ptt$ at the parton and particle levels. In general, measurements are found to be in fair agreement with predictions within the uncertainties.

The parton-level results are also compared to the following perturbative QCD calculations:

\begin{itemize}
\item An approximate NNLO calculation based on QCD threshold expansions beyond the leading-logarithmic approximation using the CT14nnlo PDF set~\cite{Guzzi:2014wia}.
\item An approximate next-to-NNLO (N$^{3}$LO) calculation performed with the resummation of soft-gluon contributions in the double-differential cross section at NNLL accuracy in momentum space using the MMHT2014 PDF set ~\cite{Kidonakis:2014pja,Harland-Lang:2014zoa}.
\item An improved NNLL QCD calculation (NLO+NNLL')~\cite{Pecjak:2016nee} with simultaneous resummation of soft and small-mass logarithms to NNLL accuracy, matched with both the standard soft-gluon resummation at NNLL accuracy and the fixed-order calculation at NLO accuracy, using the MTSW2008nnlo PDF set.
\item A full NNLO calculation based on the NNPDF3.0 PDF set~\cite{Czakon:2015owf}.
\end{itemize}

\begin{figure}
\centering
\includegraphics[width=\cmsFigWidth]{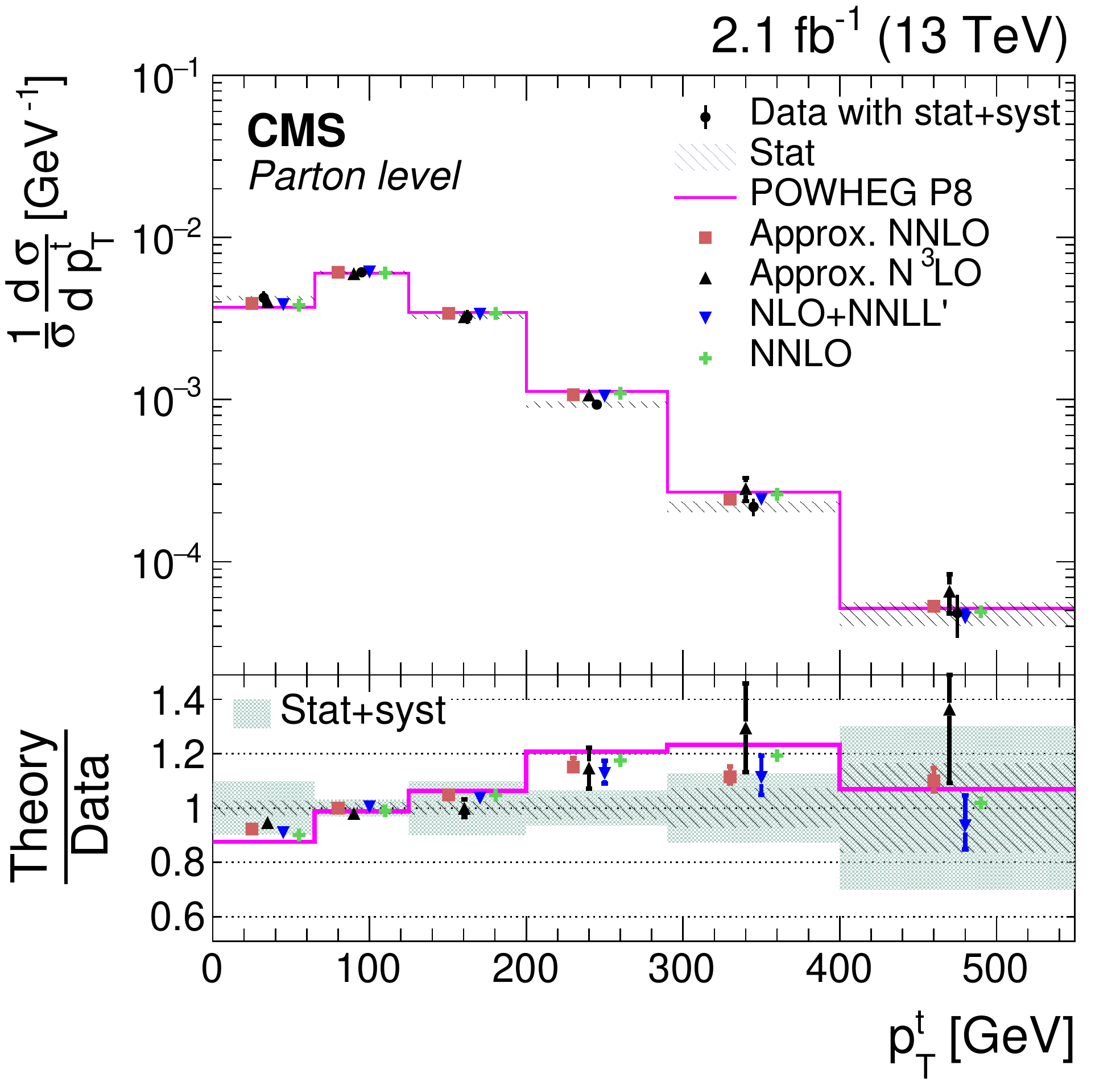}
\includegraphics[width=\cmsFigWidth]{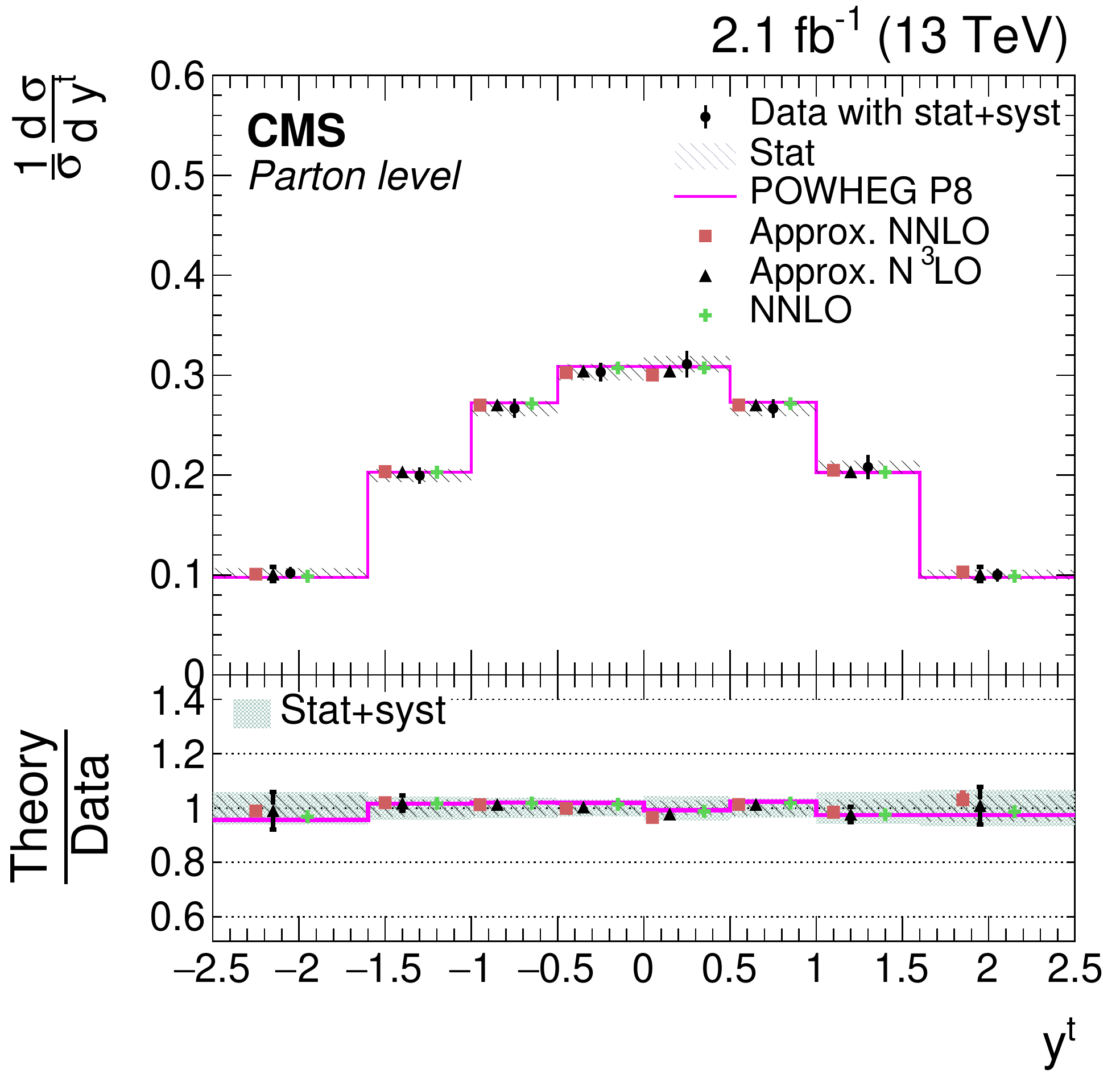}
\caption{Normalized differential \ttbar cross sections as a function of top quark \pt (left) and top quark rapidity (right), measured at the parton level in the full phase space and combining the distributions for top quarks and antiquarks. The vertical bars on the data points indicate the total (combined statistical and systematic) uncertainties, while the hatched band shows the statistical uncertainty. The measurements are compared to different perturbative QCD calculations of an approximate NNLO~\cite{Guzzi:2014wia}, an approximate next-to-NNLO (N$^{3}$LO)~\cite{Kidonakis:2014pja}, an improved NLO+NNLL (NLO+NNLL')~\cite{Pecjak:2016nee}, and  a full NNLO~\cite{Czakon:2015owf}. The lower panel gives the ratio of the theoretical predictions to the data.}
\label{fig:parton_level_3}
\vspace{10pt}
\end{figure}

\begin{figure}
\centering
\includegraphics[width=\cmsFigWidth]{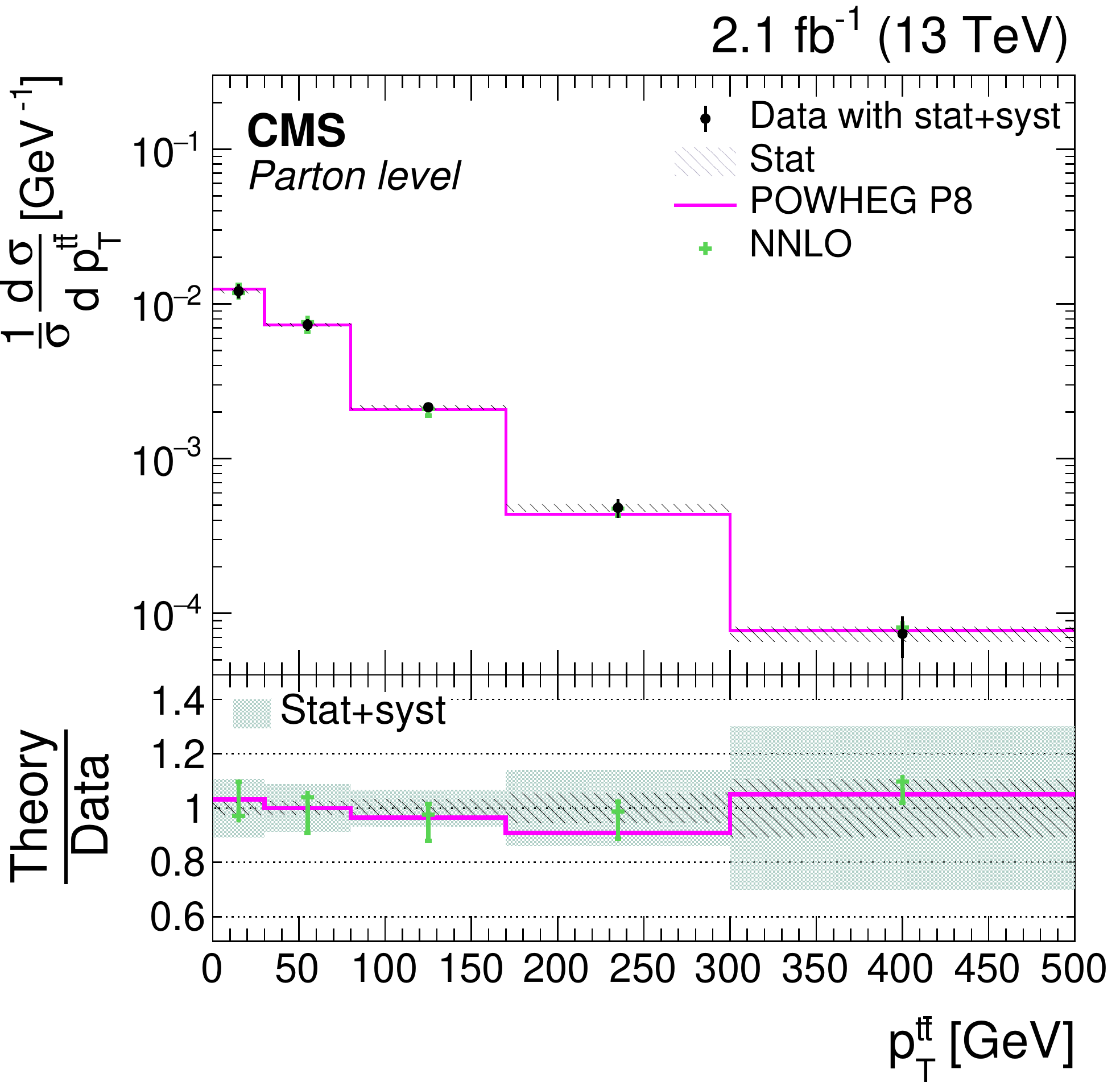}
\includegraphics[width=\cmsFigWidth]{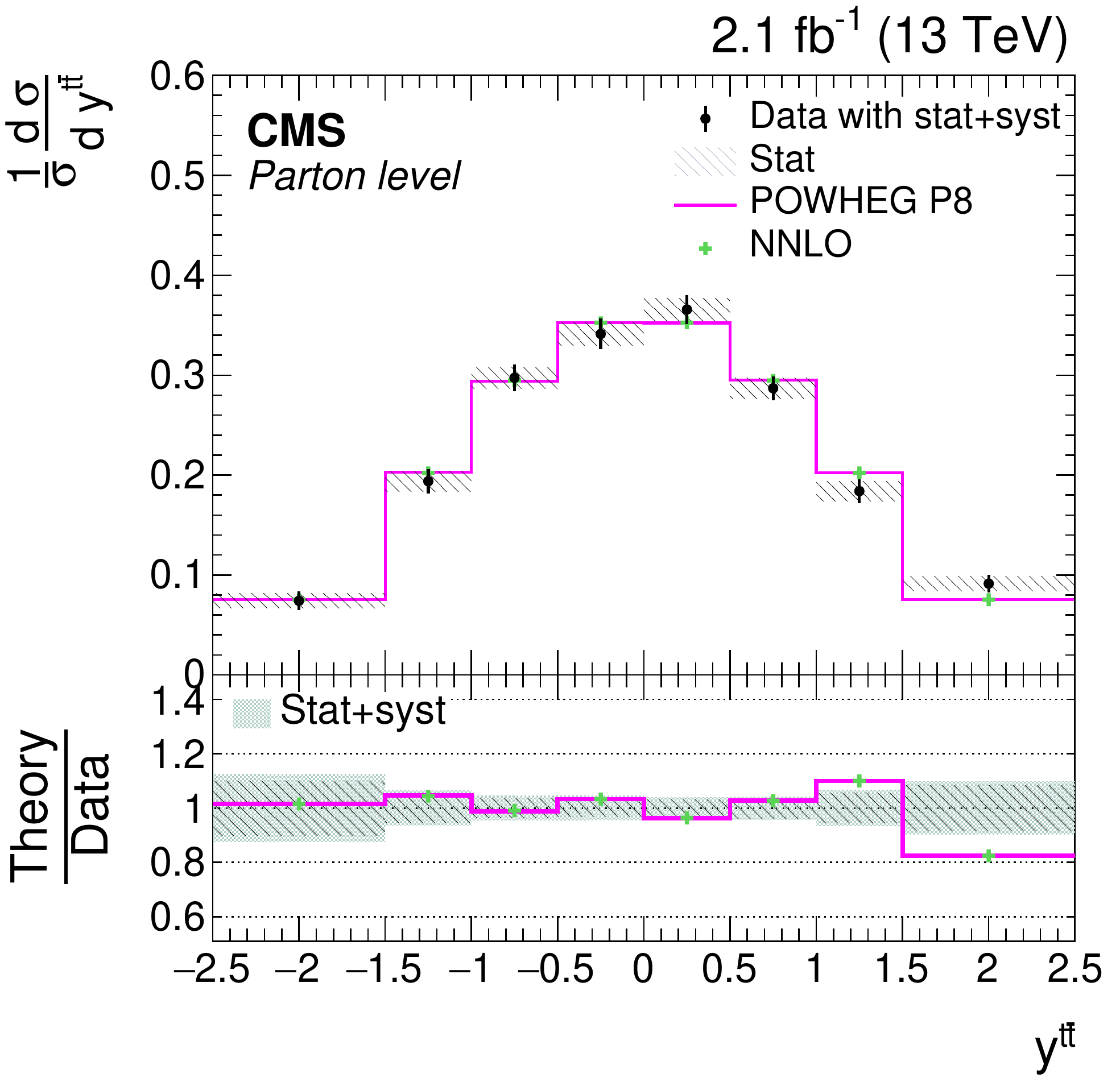}
\\ \vspace{0.2cm}
\includegraphics[width=\cmsFigWidth]{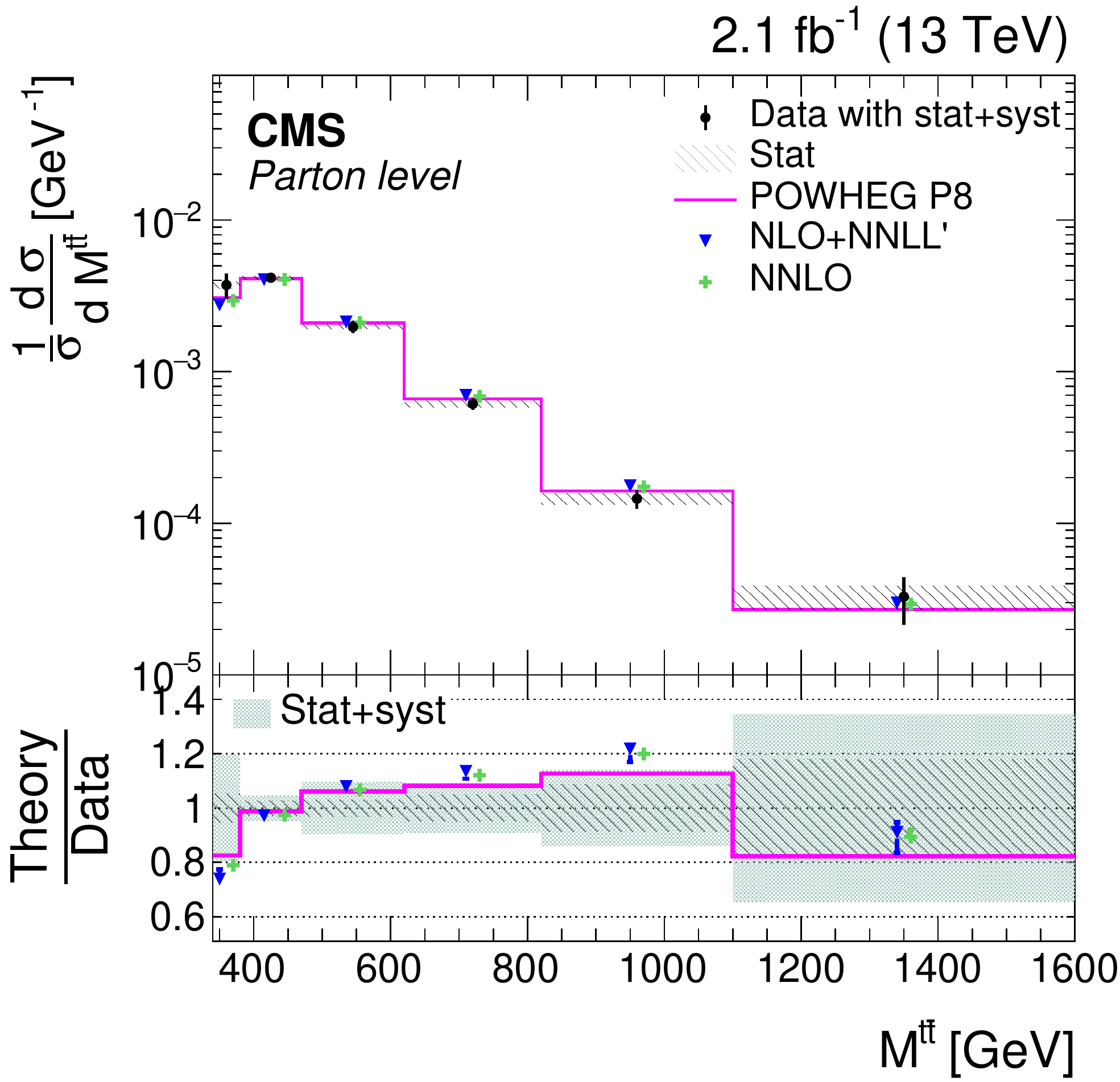}
\caption{Normalized differential \ttbar cross sections as a function of $\pttt$ (upper left), $\ytt$ (upper right), and $\mtt$ (lower) for the top quarks or antiquarks, measured at parton level in the full phase space. The vertical bars on the data points indicate the total (combined statistical and systematic) uncertainties, while the hatched band shows the statistical uncertainty. The measurements are compared to different perturbative QCD calculations of an improved NLO+NNLL (NLO+NNLL')~\cite{Pecjak:2016nee} and  a full NNLO~\cite{Czakon:2015owf}. The lower panel gives the ratio of the theoretical predictions to the data.}
\label{fig:parton_level_4}
\vspace{10pt}
\end{figure}

The measurements and the perturbative QCD predictions are shown in Figs.~\ref{fig:parton_level_3} and \ref{fig:parton_level_4}. Table~\ref{tab:chi2_gen3} gives the $\chi^{2}/$dof and the corresponding p-values for the agreement between the measurements and QCD calculations. The normalized differential \ttbar cross sections as a function of the $\yt$, $\ytt$, and $\pttt$ are found to be in good agreement with the different predictions considered. We observe some tension between the data and the NNLO predictions for other variables such as the $\ptt$ and $\mtt$.

\begin{table}[htb]
\centering
\topcaption{The $\chi^{2}/$dof and p-values for the comparison of the measured normalized \ttbar differential cross sections with different model predictions at the particle level for each of the kinematic variables.}\label{tab:chi2_gen1}
\renewcommand{\arraystretch}{1.2}
\resizebox{\textwidth}{!}
{
\begin{tabular}{c|cc|cc|cc|cc}
& \multicolumn{2}{c|}{\POWHEG} & \multicolumn{2}{c|}{\aMCatNLO} & \multicolumn{2}{c|}{\aMCatNLO} & \multicolumn{2}{c}{\POWHEG}  \\
& \multicolumn{2}{c|}{+~\PYTHIA{}8} & \multicolumn{2}{c|}{+~\PYTHIA{}8 \MLM} & \multicolumn{2}{c|}{+~\PYTHIA{}8 \FXFX} & \multicolumn{2}{c}{+~\HERWIGpp}  \\ \hline
Variable & $\x\chi^{2}/$dof  & p-value & $\x\chi^{2}/$dof  & p-value & $\x\chi^{2}/$dof  & p-value & $\x\chi^{2}/$dof  & p-value \\ \hline
$\ptlep$ &  63.4/6 &  $<$0.01\z & 79.5/6 &  $<$0.01\z & 44.1/6 &  $<$0.01\z & 20.2/6 &  $<$0.01\z  \\
$\ptjet$ &  32.4/4 &  $<$0.01\z & 60.0/4 &  $<$0.01\z & 10.8/4 &  \x0.029 & 11.1/4 &  0.03 \\
$\ptt$ &  57.2/5 &  $<$0.01\z & 77.7/5 &  $<$0.01\z & 31.6/5 &  $<$0.01\z & \x4.2/5 &  0.53 \\
$\yt$ &  \x5.1/7 &  0.65 & \x4.7/7 &  0.69 & \x3.7/7 &  0.81 & \x4.9/7 &  0.67 \\
$\pttt$ &  \x2.6/4 &  0.62 & \x7.1/4 &  0.13 & 13.1/4 &  0.01 & \x9.5/4 &  0.05 \\
$\ytt$ &   \x8.6/7 &  0.28 & 12.3/7 &  0.09 & \x8.8/7 &  0.26 & 10.0/7 &  0.19 \\
$\mtt$ & 16.9/4 &  $<$0.01\z & 16.5/4 &  $<$0.01\z & \x5.3/4 &  0.26 & 14.2/4 &  $<$0.01\z  \\
$\dphtt$ &  14.7/3 &  $<$0.01\z & \x1.1/3 &  0.79 & \x1.3/3 &  0.74 & \x9.7/3 &  0.02
\end{tabular}
}
\end{table}

\begin{table}[htb]
\centering
\topcaption{The $\chi^{2}/$dof and p-values for the comparison of the measured normalized \ttbar differential cross sections with different model predictions at the parton level for each of the kinematic variables.}\label{tab:chi2_gen2}
\renewcommand{\arraystretch}{1.2}
\resizebox{\textwidth}{!}
{
\begin{tabular}{c|cc|cc|cc|cc}
& \multicolumn{2}{c|}{\POWHEG} & \multicolumn{2}{c|}{\aMCatNLO} & \multicolumn{2}{c|}{\aMCatNLO} & \multicolumn{2}{c}{\POWHEG}  \\
& \multicolumn{2}{c|}{+~\PYTHIA{}8} & \multicolumn{2}{c|}{+~\PYTHIA{}8 \MLM} & \multicolumn{2}{c|}{+~\PYTHIA{}8 \FXFX} & \multicolumn{2}{c}{+~\HERWIGpp}  \\ \hline
Variable & $\x\chi^{2}/$dof  & p-value & $\x\chi^{2}/$dof  & p-value & $\x\chi^{2}/$dof  & p-value & $\x\chi^{2}/$dof  & p-value \\  \hline
$\ptt$ &  67.6/5 &  $<$0.01\z & 99.1/5 &  $<$0.01\z & 49.4/5 &  $<$0.01\z & 19.0/5 &  $<$0.01\z  \\
$\yt$ &  \x4.4/7 &  0.73 & \x5.1/7 &  0.65 & \x5.4/7 &  0.61 & \x5.3/7 &  0.63 \\
$\pttt$ &  \x4.4/4 &  0.35 & 24.1/4 &  $<$0.01\z & 38.7/4 &  $<$0.01\z & 19.2/4 &  $<$0.01\z  \\
$\ytt$ &  \x7.7/7 &  0.36 & \x9.2/7 &  0.24 & \x9.3/7 &  0.23 & \x8.0/7 &  0.33 \\
$\mtt$ & 21.2/5 &  $<$0.01\z & \x6.5/5 &  0.26 & \x4.3/5 &  0.51 & \x1.6/5 &  0.90 \\
$\dphtt$ & 22.3/3 &  $<$0.01\z & \x1.7/3 &  0.65 & \x3.9/3 &  0.28 & 27.9/3 &  $<$0.01\z  \\
\end{tabular}
}
\end{table}

\begin{table}[htb]
\centering
\topcaption{The $\chi^{2}/$dof and p-values for the comparison of the measured normalized \ttbar differential cross sections with published perturbative QCD calculations.}\label{tab:chi2_gen3}
\renewcommand{\arraystretch}{1.2}
\resizebox{\textwidth}{!}
{
\begin{tabular}{c|cc|cc|cc|cc}
& \multicolumn{2}{c|}{Approx. NNLO~\cite{Guzzi:2014wia}} & \multicolumn{2}{c|}{Approx. N$^{3}$LO~\cite{Kidonakis:2014pja}} & \multicolumn{2}{c|}{NLO+NNLL'~\cite{Pecjak:2016nee}} & \multicolumn{2}{c}{NNLO~\cite{Czakon:2015owf}} \\ \hline
Variable & $\x\chi^{2}/$dof  & p-value & $\x\chi^{2}/$dof  & p-value & $\x\chi^{2}/$dof  & p-value & $\x\chi^{2}/$dof  & p-value \\ \hline
$\ptt$ & 27.9/5 &  $<$0.01\z & 43.8/5 &  $<$0.01\z & 24.1/5 &  $<$0.01\z & 44.8/5 &  $<$0.01\z  \\
$\yt$ & \x4.2/7 &  0.76 & \x3.75/7 &  0.81 &  & & \x3.8/7 &  0.80 \\
$\pttt$ &  & &  & &  & & \x4.0/4 &  0.40 \\
$\ytt$ &  & &  & &  & & \x7.6/7 &  0.37 \\
$\mtt$ &  & &  & & 68.3/5 &  $<$0.01\z & 47.6/5 &  $<$0.01\z  \\
\end{tabular}
}
\end{table}

\section{Summary}
\label{sec:summary}

The normalized differential cross sections for top quark pair production have been presented by the CMS experiment in the dilepton decay channel in pp collisions at $\sqrt{s} = 13\TeV$ with data corresponding to an integrated luminosity of 2.1\fbinv. The differential cross sections are measured as a function of several kinematic variables at particle level in a visible phase space corresponding to the detector acceptance and at parton level in the full phase space. The measurements are compared to the predictions from Monte Carlo simulations and calculations in perturbative quantum chromodynamics. In general, the measurements are in fairly good agreement with predictions. We confirm that the top quark \pt spectrum in data is softer than the Monte Carlo predictions at both particle and parton levels, as reported by the ATLAS and CMS experiments. The present results are in agreement with the earlier ATLAS and CMS measurements. We also find that the measurements are in better agreement with calculations within quantum chromodynamics up to next-to-next-to-leading-order accuracy at the parton level compared to previous next-to-leading-order predictions.

\clearpage

\begin{acknowledgments}

\hyphenation{Bundes-ministerium Forschungs-gemeinschaft Forschungs-zentren Rachada-pisek} We congratulate our colleagues in the CERN accelerator departments for the excellent performance of the LHC and thank the technical and administrative staffs at CERN and at other CMS institutes for their contributions to the success of the CMS effort. In addition, we gratefully acknowledge the computing centres and personnel of the Worldwide LHC Computing Grid for delivering so effectively the computing infrastructure essential to our analyses. Finally, we acknowledge the enduring support for the construction and operation of the LHC and the CMS detector provided by the following funding agencies: the Austrian Federal Ministry of Science, Research and Economy and the Austrian Science Fund; the Belgian Fonds de la Recherche Scientifique, and Fonds voor Wetenschappelijk Onderzoek; the Brazilian Funding Agencies (CNPq, CAPES, FAPERJ, and FAPESP); the Bulgarian Ministry of Education and Science; CERN; the Chinese Academy of Sciences, Ministry of Science and Technology, and National Natural Science Foundation of China; the Colombian Funding Agency (COLCIENCIAS); the Croatian Ministry of Science, Education and Sport, and the Croatian Science Foundation; the Research Promotion Foundation, Cyprus; the Secretariat for Higher Education, Science, Technology and Innovation, Ecuador; the Ministry of Education and Research, Estonian Research Council via IUT23-4 and IUT23-6 and European Regional Development Fund, Estonia; the Academy of Finland, Finnish Ministry of Education and Culture, and Helsinki Institute of Physics; the Institut National de Physique Nucl\'eaire et de Physique des Particules~/~CNRS, and Commissariat \`a l'\'Energie Atomique et aux \'Energies Alternatives~/~CEA, France; the Bundesministerium f\"ur Bildung und Forschung, Deutsche Forschungsgemeinschaft, and Helmholtz-Gemeinschaft Deutscher Forschungszentren, Germany; the General Secretariat for Research and Technology, Greece; the National Scientific Research Foundation, and National Innovation Office, Hungary; the Department of Atomic Energy and the Department of Science and Technology, India; the Institute for Studies in Theoretical Physics and Mathematics, Iran; the Science Foundation, Ireland; the Istituto Nazionale di Fisica Nucleare, Italy; the Ministry of Science, ICT and Future Planning, and National Research Foundation (NRF), Republic of Korea; the Lithuanian Academy of Sciences; the Ministry of Education, and University of Malaya (Malaysia); the Mexican Funding Agencies (BUAP, CINVESTAV, CONACYT, LNS, SEP, and UASLP-FAI); the Ministry of Business, Innovation and Employment, New Zealand; the Pakistan Atomic Energy Commission; the Ministry of Science and Higher Education and the National Science Centre, Poland; the Funda\c{c}\~ao para a Ci\^encia e a Tecnologia, Portugal; JINR, Dubna; the Ministry of Education and Science of the Russian Federation, the Federal Agency of Atomic Energy of the Russian Federation, Russian Academy of Sciences, the Russian Foundation for Basic Research and the Russian Competitiveness Program of NRNU ``MEPhI"; the Ministry of Education, Science and Technological Development of Serbia; the Secretar\'{\i}a de Estado de Investigaci\'on, Desarrollo e Innovaci\'on, Programa Consolider-Ingenio 2010, Plan de Ciencia, Tecnolog\'{i}a e Innovaci\'on 2013-2017 del Principado de Asturias and Fondo Europeo de Desarrollo Regional, Spain; the Swiss Funding Agencies (ETH Board, ETH Zurich, PSI, SNF, UniZH, Canton Zurich, and SER); the Ministry of Science and Technology, Taipei; the Thailand Center of Excellence in Physics, the Institute for the Promotion of Teaching Science and Technology of Thailand, Special Task Force for Activating Research and the National Science and Technology Development Agency of Thailand; the Scientific and Technical Research Council of Turkey, and Turkish Atomic Energy Authority; the National Academy of Sciences of Ukraine, and State Fund for Fundamental Researches, Ukraine; the Science and Technology Facilities Council, UK; the US Department of Energy, and the US National Science Foundation.

Individuals have received support from the Marie-Curie program and the European Research Council and Horizon 2020 Grant, contract No. 675440 (European Union); the Leventis Foundation; the A. P. Sloan Foundation; the Alexander von Humboldt Foundation; the Belgian Federal Science Policy Office; the Fonds pour la Formation \`a la Recherche dans l'Industrie et dans l'Agriculture (FRIA-Belgium); the Agentschap voor Innovatie door Wetenschap en Technologie (IWT-Belgium); the Ministry of Education, Youth and Sports (MEYS) of the Czech Republic; the Council of Scientific and Industrial Research, India; the HOMING PLUS program of the Foundation for Polish Science, cofinanced from European Union, Regional Development Fund, the Mobility Plus program of the Ministry of Science and Higher Education, the National Science Center (Poland), contracts Harmonia 2014/14/M/ST2/00428, Opus 2014/13/B/ST2/02543, 2014/15/B/ST2/03998, and 2015/19/B/ST2/02861, Sonata-bis 2012/07/E/ST2/01406; the National Priorities Research Program by Qatar National Research Fund; the Programa Clar\'in-COFUND del Principado de Asturias; the Thalis and Aristeia programs cofinanced by EU-ESF and the Greek NSRF; the Rachadapisek Sompot Fund for Postdoctoral Fellowship, Chulalongkorn University and the Chulalongkorn Academic into Its 2nd Century Project Advancement Project (Thailand); and the Welch Foundation, contract C-1845.

\end{acknowledgments}

\bibliography{auto_generated}

\clearpage

\appendix

\begin{appendix}

\clearpage

\section{Tables of differential \texorpdfstring{\ttbar}{t-tbar} cross sections at the particle level}

\begin{table}[htb]
\centering
\topcaption{Normalized differential \ttbar cross sections with statistical and systematic uncertainties at the particle level as a function of $\ptlep$. The factor given in the last column applies to the values of the normalized cross section and the statistical and systematic uncertainties in that row.}
\label{tab:tab_lpt_c}
\renewcommand{\arraystretch}{1.2}
\begin{tabular}{c |c | c | c | c}
$\ptlep$ [\GeVns{}] &  (1/$\sigma$)($\rd \sigma$/$\rd \ptlep$) & stat & syst &  factor \\ \hline
$[20, 30]$ & 2.00 & 0.04 & 0.03 &  ${\times}10^{-2}$ \\
$[30, 40]$ & 1.84 & 0.04 & 0.03 &  ${\times}10^{-2}$ \\
$[40, 60]$ & 1.38 & 0.02 & 0.01 &  ${\times}10^{-2}$ \\
$[60, 80]$ & 8.12 & 0.17 & 0.11 &  ${\times}10^{-3}$ \\
$\x[80, 120]$ & 3.12 & 0.07 & 0.09 &  ${\times}10^{-3}$ \\
$[120, 180]$ & 6.79 & 0.29 & 0.25 &  ${\times}10^{-4}$ \\
$[180, 400]$ & 5.01 & 0.44 & 0.33 &  ${\times}10^{-5}$ \\
\end{tabular}
\end{table}

\begin{table}[htb]
\centering
\topcaption{Normalized differential \ttbar cross sections with statistical and systematic uncertainties at the particle level as a function of $\ptjet$. The factor given in the last column applies to the values of the normalized cross section and the statistical and systematic uncertainties in that row.}
\label{tab:tab_jpt_c}
\renewcommand{\arraystretch}{1.2}
\begin{tabular}{c |c | c | c | c}
$\ptjet$ [\GeVns{}] &  (1/$\sigma$)($\rd \sigma$/$\rd \ptjet$) & stat & syst &  factor \\ \hline
$[30, 50]$ & 1.42 & 0.03 & 0.07 &  ${\times}10^{-2}$ \\
$[50, 80]$ & 1.12 & 0.02 & 0.02 &  ${\times}10^{-2}$ \\
$\x[80, 130]$ & 5.24 & 0.11 & 0.18 &  ${\times}10^{-3}$ \\
$[130, 210]$ & 1.18 & 0.04 & 0.06 &  ${\times}10^{-3}$ \\
$[210, 500]$ & 8.2 & 0.71 & 2.1 &  ${\times}10^{-5}$ \\
\end{tabular}
\end{table}

\begin{table}[htb]
\centering
\topcaption{Normalized differential \ttbar cross sections with statistical and systematic uncertainties at the particle level as a function of $\ptt$. The factor given in the last column applies to the values of the normalized cross section and the statistical and systematic uncertainties in that row.}
\label{tab:tab_tpt_c}
\renewcommand{\arraystretch}{1.2}
\begin{tabular}{c |c | c | c | c}
$\ptt$ [\GeVns{}] &  (1/$\sigma$)($\rd \sigma$/$\rd \ptt$) & stat & syst &  factor \\ \hline
$\x[0, 65]$ & 4.14 & 0.12 & 0.13 &  ${\times}10^{-3}$ \\
$\x[65, 125]$ & 5.73 & 0.16 & 0.23 &  ${\times}10^{-3}$ \\
$[125, 200]$ & 3.20 & 0.10 & 0.13 &  ${\times}10^{-3}$ \\
$[200, 290]$ & 1.08 & 0.05 & 0.09 &  ${\times}10^{-3}$ \\
$[290, 400]$ & 3.42 & 0.27 & 0.54 &  ${\times}10^{-4}$ \\
$[400, 550]$ & 7.9 & 1.5 & 2.6 &  ${\times}10^{-5}$ \\
\end{tabular}
\end{table}

\begin{table}[htb]
\centering
\topcaption{Normalized differential \ttbar cross sections with statistical and systematic uncertainties at the particle level as a function of $\yt$. The factor given in the last column applies to the values of the normalized cross section and the statistical and systematic uncertainties in that row.}
\label{tab:tab_ty_c}
\renewcommand{\arraystretch}{1.2}
\begin{tabular}{c |c | c | c | c}
$\yt$  &  (1/$\sigma$)($\rd \sigma$/$\rd \yt$) & stat & syst &  factor \\ \hline
$[-2.5, -1.6]$ & 5.80 & 0.34 & 0.23 &  ${\times}10^{-2}$ \\
$[-1.6, -1.0]$ & 2.02 & 0.08 & 0.08 &  ${\times}10^{-1}$ \\
$[-1.0, -0.5]$ & 2.95 & 0.10 & 0.07 &  ${\times}10^{-1}$ \\
$[-0.5, 0.0]\w$ & 3.45 & 0.11 & 0.05 &  ${\times}10^{-1}$ \\
$[0.0, 0.5]$ & 3.57 & 0.11 & 0.11 &  ${\times}10^{-1}$ \\
$[0.5, 1.0]$ & 2.98 & 0.10 & 0.05 &  ${\times}10^{-1}$ \\
$[1.0, 1.6]$ & 2.12 & 0.08 & 0.08 &  ${\times}10^{-1}$ \\
$[1.6, 2.5]$ & 5.71 & 0.34 & 0.26 &  ${\times}10^{-2}$ \\
\end{tabular}
\end{table}

\begin{table}[htb]
\centering
\topcaption{Normalized differential \ttbar cross sections with statistical and systematic uncertainties at the particle level as a function of $\pt^{\ttbar}$. The factor given in the last column applies to the values of the normalized cross section and the statistical and systematic uncertainties in that row.}
\label{tab:tab_ttpt_c}
\renewcommand{\arraystretch}{1.2}
\begin{tabular}{c |c | c | c | c}
$\pttt$ [\GeVns{}] &  (1/$\sigma$)($\rd \sigma$/$\rd \pttt$) & stat & syst &  factor \\ \hline
$\x[0, 30]$ & 1.01 & 0.03 & 0.14 &  ${\times}10^{-2}$ \\
$[30, 80]$ & 8.16 & 0.26 & 0.65 &  ${\times}10^{-3}$ \\
$\x[80, 170]$ & 2.34 & 0.10 & 0.17 &  ${\times}10^{-3}$ \\
$[170, 300]$ & 4.81 & 0.39 & 0.72 &  ${\times}10^{-4}$ \\
$[300, 500]$ & 7.6 & 1.3 & 2.6 &  ${\times}10^{-5}$ \\
\end{tabular}
\end{table}

\begin{table}[htb]
\centering
\topcaption{Normalized differential \ttbar cross sections with statistical and systematic uncertainties at the particle level as a function of $\ytt$. The factor given in the last column applies to the values of the normalized cross section and the statistical and systematic uncertainties in that row.}
\label{tab:tab_tty_c}
\renewcommand{\arraystretch}{1.2}
\begin{tabular}{c |c | c | c | c}
$\ytt$ &  (1/$\sigma$)($\rd \sigma$/$\rd \ytt$) & stat & syst &  factor \\ \hline
$[-2.5, -1.5]$ & 2.57 & 0.32 & 0.19 &  ${\times}10^{-2}$ \\
$[-1.5, -1.0]$ & 1.68 & 0.10 & 0.07 &  ${\times}10^{-1}$ \\
$[-1.0, -0.5]$ & 3.37 & 0.14 & 0.04 &  ${\times}10^{-1}$ \\
$[-0.5, 0.0]\w$ & 4.30 & 0.16 & 0.11 &  ${\times}10^{-1}$ \\
$[0.0, 0.5]$ & 4.60 & 0.16 & 0.06 &  ${\times}10^{-1}$ \\
$[0.5, 1.0]$ & 3.28 & 0.14 & 0.08 &  ${\times}10^{-1}$ \\
$[1.0, 1.5]$ & 1.58 & 0.10 & 0.07 &  ${\times}10^{-1}$ \\
$[1.5, 2.5]$ & 3.35 & 0.33 & 0.20 &  ${\times}10^{-2}$ \\
\end{tabular}
\end{table}

\begin{table}[htb]
\centering
\topcaption{Normalized differential \ttbar cross sections with statistical and systematic uncertainties at the particle level as a function of $\mtt$. The factor given in the last column applies to the values of the normalized cross section and the statistical and systematic uncertainties in that row.}
\label{tab:tab_ttm_c}
\renewcommand{\arraystretch}{1.2}
\begin{tabular}{c |c | c | c | c}
$\mtt$ [\GeVns{}] &  (1/$\sigma$)($\rd \sigma$/$\rd \mtt$)& stat & syst &  factor \\ \hline
$[300, 400]$ & 3.07 & 0.13 & 0.12 &  ${\times}10^{-3}$ \\
$[400, 500]$ & 3.07 & 0.15 & 0.20 &  ${\times}10^{-3}$ \\
$[500, 650]$ & 1.44 & 0.08 & 0.08 &  ${\times}10^{-3}$ \\
$\x[650, 1000]$ & 3.85 & 0.26 & 0.80 &  ${\times}10^{-4}$ \\
$[1000, 1600]$ & 5.9 & 0.90 & 1.6 &  ${\times}10^{-5}$ \\
\end{tabular}
\end{table}

\begin{table}[ht]
\centering
\topcaption{Normalized differential \ttbar cross sections with statistical and systematic uncertainties at the particle level as a function of $\dphtt$. The factor given in the last column applies to the values of the normalized cross section and the statistical and systematic uncertainties in that row.}
\label{tab:tab_ttdphi_c}
\renewcommand{\arraystretch}{1.2}
\begin{tabular}{c |c | c | c | c}
$\dphtt$ [rad] &  (1/$\sigma$)($\rd \sigma$/$\rd \dphtt$ ) & stat & syst &  factor \\ \hline
$\y\x\x[0, 1.57]$ & 6.79 & 0.60 & 1.04 &  ${\times}10^{-2}$ \\
$[1.57, 2.61]$ & 2.26 & 0.14 & 0.26 &  ${\times}10^{-1}$ \\
$\x[2.61, 3.016]$ & 9.52 & 0.44 & 0.71 &  ${\times}10^{-1}$ \\
$[3.016, 3.142]$ & 2.2 & 0.10 & 0.41 &  ${\times}1\phantom{0^{-0}}$ \\
\end{tabular}
\end{table}

\clearpage

\section{Tables of differential cross section at the parton level}

\begin{table}[htb]
\centering
\topcaption{Normalized differential \ttbar cross sections with statistical and systematic uncertainties at the parton level as a function of $\pt^{\PQt}$. The factor given in the last column applies to the values of the normalized cross section and the statistical and systematic uncertainties in that row.}
\label{tab:tab_tpt_p}
\renewcommand{\arraystretch}{1.2}
\begin{tabular}{c |c | c | c | c}
$\ptt$ [\GeVns{}] &  (1/$\sigma$)($\rd \sigma$/$\rd \ptt$) & stat & syst &  factor \\ \hline
$\x[0, 65]$ & 4.24 & 0.11 & 0.40 &  ${\times}10^{-3}$ \\
$\x[65, 125]$ & 6.10 & 0.13 & 0.14 &  ${\times}10^{-3}$ \\
$[125, 200]$ & 3.25 & 0.08 & 0.31 &  ${\times}10^{-3}$ \\
$[200, 290]$ & 9.31 & 0.37 & 0.47 &  ${\times}10^{-4}$ \\
$[290, 400]$ & 2.18 & 0.16 & 0.22 &  ${\times}10^{-4}$ \\
$[400, 550]$ & 4.8 & 0.79 & 1.2 &  ${\times}10^{-5}$ \\
\end{tabular}
\end{table}

\begin{table}[htb]
\centering
\topcaption{Normalized differential \ttbar cross sections with statistical and systematic uncertainties at the parton level as a function of $\yt$. The factor given in the last column applies to the values of the normalized cross section and the statistical and systematic uncertainties in that row.}
\label{tab:tab_ty_p}
\renewcommand{\arraystretch}{1.2}
\begin{tabular}{c |c | c | c | c}
$\yt$ &  (1/$\sigma$)($\rd \sigma$/$\rd \yt$) & stat & syst &  factor \\ \hline
$[-2.5, -1.6]$ & 1.02 & 0.05 & 0.03 &  ${\times}10^{-1}$ \\
$[-1.6, -1.0]$ & 1.99 & 0.06 & 0.05 &  ${\times}10^{-1}$ \\
$[-1.0, -0.5]$ & 2.67 & 0.08 & 0.06 &  ${\times}10^{-1}$ \\
$[-0.5, 0.0]\w$ & 3.03 & 0.08 & 0.04 &  ${\times}10^{-1}$ \\
$[0.0, 0.5]$ & 3.11 & 0.08 & 0.11 &  ${\times}10^{-1}$ \\
$[0.5, 1.0]$ & 2.67 & 0.08 & 0.05 &  ${\times}10^{-1}$ \\
$[1.0, 1.6]$ & 2.08 & 0.06 & 0.10 &  ${\times}10^{-1}$ \\
$[1.6, 2.5]$ & 1.00 & 0.05 & 0.05 &  ${\times}10^{-1}$ \\
\end{tabular}
\end{table}

\begin{table}[htb]
\centering
\topcaption{Normalized differential \ttbar cross sections with statistical and systematic uncertainties at the parton level as a function of $\pt^{\ttbar}$. The factor given in the last column applies to the values of the normalized cross section and the statistical and systematic uncertainties in that row.}
\label{tab:tab_ttpt_p}
\renewcommand{\arraystretch}{1.2}
\begin{tabular}{c |c | c | c | c}
$\pttt$ [\GeVns{}] &  (1/$\sigma$)($\rd \sigma$/$\rd \pttt$) & stat & syst &  factor \\ \hline
$\x[0, 30]$ & 1.21 & 0.03 & 0.13 &  ${\times}10^{-2}$ \\
$[30, 80]$ & 7.32 & 0.18 & 0.61 &  ${\times}10^{-3}$ \\
$\x[80, 170]$ & 2.15 & 0.07 & 0.13 &  ${\times}10^{-3}$ \\
$[170, 300]$ & 4.81 & 0.27 & 0.62 &  ${\times}10^{-4}$ \\
$[300, 500]$ & 7.4 & 0.79 & 2.1 &  ${\times}10^{-5}$ \\
\end{tabular}
\end{table}

\begin{table}[htb]
\centering
\topcaption{Normalized differential \ttbar cross sections with statistical and systematic uncertainties at the parton level as a function of $\ytt$. The factor given in the last column applies to the values of the normalized cross section and the statistical and systematic uncertainties in that row.}
\label{tab:tab_tty_p}
\renewcommand{\arraystretch}{1.2}
\begin{tabular}{c |c | c | c | c}
$\ytt$ &  (1/$\sigma$)($\rd \sigma$/$\rd \ytt$) &  stat & syst & factor \\ \hline
$[-2.5, -1.5]$ & 7.42 & 0.75 & 0.54 &  ${\times}10^{-2}$ \\
$[-1.5, -1.0]$ & 1.94 & 0.10 & 0.07 &  ${\times}10^{-1}$ \\
$[-1.0, -0.5]$ & 2.97 & 0.11 & 0.08 &  ${\times}10^{-1}$ \\
$[-0.5, 0.0]\w$ & 3.41 & 0.11 & 0.10 &  ${\times}10^{-1}$ \\
$[0.0, 0.5]$ & 3.66 & 0.11 & 0.09 &  ${\times}10^{-1}$ \\
$[0.5, 1.0]$ & 2.87 & 0.11 & 0.05 &  ${\times}10^{-1}$ \\
$[1.0, 1.5]$ & 1.84 & 0.10 & 0.07 &  ${\times}10^{-1}$ \\
$[1.5, 2.5]$ & 9.14 & 0.76 & 0.45 &  ${\times}10^{-2}$ \\
\end{tabular}
\end{table}

\begin{table}[htb]
\centering
\topcaption{Normalized differential \ttbar cross sections with statistical and systematic uncertainties at the parton level  as a function of $\mtt$. The factor given in the last column applies to the values of the normalized cross section and the statistical and systematic uncertainties in that row.}
\label{tab:tab_ttm_p}
\renewcommand{\arraystretch}{1.2}
\begin{tabular}{c |c | c | c | c}
$\mtt$ [\GeVns{}] &  (1/$\sigma$)($\rd \sigma$/$\rd \mtt$) & stat & syst &  factor \\ \hline
$[340, 380]$ & 3.73 & 0.20 & 0.70 &  ${\times}10^{-3}$ \\
$[380, 470]$ & 4.16 & 0.11 & 0.15 &  ${\times}10^{-3}$ \\
$[470, 620]$ & 1.97 & 0.06 & 0.18 &  ${\times}10^{-3}$ \\
$[620, 820]$ & 6.14 & 0.30 & 0.48 &  ${\times}10^{-4}$ \\
$\x[820, 1100]$ & 1.45 & 0.13 & 0.15 &  ${\times}10^{-4}$ \\
$[1100, 1600]$ & 3.28 & 0.59 & 0.97 &  ${\times}10^{-5}$ \\
\end{tabular}
\end{table}

\begin{table}[htb]
\centering
\topcaption{Normalized differential \ttbar cross sections with statistical and systematic uncertainties at the parton level as a function of $\dphtt$. The factor given in the last column applies to the values of the normalized cross section and the statistical and systematic uncertainties in that row.}
\label{tab:tab_ttdphi_p}
\renewcommand{\arraystretch}{1.2}
\begin{tabular}{c |c | c | c | c}
$\dphtt$ [rad] &  (1/$\sigma$)($\rd \sigma$/$\rd \dphtt$)& stat & syst &  factor \\ \hline
$\x\x\y[0, 1.57]$ & 7.02 & 0.48 & 0.92 &  ${\times}10^{-2}$ \\
$[1.57, 2.61]$ & 2.14 & 0.11 & 0.24 &  ${\times}10^{-1}$ \\
$\x[2.61, 3.016]$ & 9.30 & 0.37 & 0.53 &  ${\times}10^{-1}$ \\
$[3.016, 3.142]$ & 2.30 & 0.09 & 0.33 &  ${\times}1\phantom{0^{-0}}$ \\
\end{tabular}
\end{table}

\end{appendix}
\cleardoublepage \section{The CMS Collaboration \label{app:collab}}\begin{sloppypar}\hyphenpenalty=5000\widowpenalty=500\clubpenalty=5000\textbf{Yerevan Physics Institute,  Yerevan,  Armenia}\\*[0pt]
A.M.~Sirunyan, A.~Tumasyan
\vskip\cmsinstskip
\textbf{Institut f\"{u}r Hochenergiephysik,  Wien,  Austria}\\*[0pt]
W.~Adam, F.~Ambrogi, E.~Asilar, T.~Bergauer, J.~Brandstetter, E.~Brondolin, M.~Dragicevic, J.~Er\"{o}, M.~Flechl, M.~Friedl, R.~Fr\"{u}hwirth\cmsAuthorMark{1}, V.M.~Ghete, J.~Grossmann, J.~Hrubec, M.~Jeitler\cmsAuthorMark{1}, A.~K\"{o}nig, N.~Krammer, I.~Kr\"{a}tschmer, D.~Liko, T.~Madlener, I.~Mikulec, E.~Pree, D.~Rabady, N.~Rad, H.~Rohringer, J.~Schieck\cmsAuthorMark{1}, R.~Sch\"{o}fbeck, M.~Spanring, D.~Spitzbart, J.~Strauss, W.~Waltenberger, J.~Wittmann, C.-E.~Wulz\cmsAuthorMark{1}, M.~Zarucki
% ...
\vskip\cmsinstskip
% ...
1:~~Also at Vienna University of Technology, Vienna, Austria\\
% ...

\end{sloppypar}
\end{document}